\documentclass[aps,twocolumn,preprintnumbers,amsmath,amssymb,superscriptaddress,longbibliography]{revtex4-1}  
\usepackage{subfigure}
\usepackage{graphicx}
\usepackage{dcolumn}
\usepackage{xcolor}

\usepackage{color}
\usepackage{amsmath}
\usepackage[breaklinks,colorlinks,bookmarks=false,citecolor=blue,linkcolor=blue,urlcolor=blue]{hyperref}
\usepackage{braket}
\usepackage[utf8x]{inputenc} 
\graphicspath{{figures/}} 

\begin{document}

\preprint{\href{https://doi.org/10.1103/PhysRevB.104.245125}{Phys. Rev. B {\bfseries 104}, 245125 (2021)}}

\title{Electronic transport properties 
and quantum localization effects 
monitored by selective functionalization in Bernal bilayer graphene}

\author{Jouda Jemaa \surname{Khabthani}}
\email{jouda.khabthani@fst.utm.tn}
\affiliation{
Laboratoire de Physique de la Mati\`ere Condens\'ee,
D\'epartement de Physique, Facult\'e des Sciences de Tunis, Universit\'e de Tunis El Manar, Campus universitaire 1060 Tunis, Tunisia}

\author{Ahmed \surname{Missaoui}}
\email{ahmed.missaoui@cyu.fr}
\affiliation{ 
Laboratoire de Physique Th\'eorique et Mod\'elisation,
CY Cergy Paris Universit\'e, CNRS, 
95302 Cergy-Pontoise, France}

\author{Didier \surname{Mayou}}
\email{didier.mayou@neel.cnrs.fr}
\affiliation{
Univ. Grenoble Alpes, Inst. NEEL, F-38042 Grenoble, France \\
CNRS, Inst. NEEL, F-38042 Grenoble, France}

\author{Guy \surname{Trambly de Laissardi\`ere}}
\email{guy.trambly@cyu.fr}
\affiliation{ 
Laboratoire de Physique Th\'eorique et Mod\'elisation,
CY Cergy Paris Universit\'e, CNRS, 
95302 Cergy-Pontoise, France}

\date{\today}

\begin{abstract}

Monitoring electronic properties of 2D materials is an essential step to open a way for applications such as electronic devices and sensors. 
From this perspective, Bernal bilayer graphene (BLG)
is a fairly simple system that offers great possibilities for tuning electronic gap and charge carriers' mobility 
by selective functionalization (adsorptions of atoms or molecules).
Here, we present a detailed numerical study of BLG electronic properties when two types of adsorption site are present simultaneously. 
We focus on realistic cases that could be realized experimentally 
with adsorbate concentration  $c$ varying from 0.25\% to 5\%.
For a given value of $c$, when the electronic doping is lower than $c$
we show that quantum effects, which are ignored in usual semi-classical calculations, strongly affect the electronic structure and the transport properties. 
A wide range of behaviors is indeed found, such as gap opening, metallic behavior or abnormal conductivity, which depend on the adsorbate positions, the $c$ value, the doping, and eventually the coupling between midgap states which can create a midgap band. 
These behaviors are understood by simple arguments based on the fact that BLG lattice is bipartite. 
We also analyze the conductivity 
at low temperature, where multiple scattering effects cannot be ignored. 
Moreover, when the Fermi energy lies in the band of midgap states, the average velocity of charge carriers cancels 
but conduction is still possible thanks to quantum fluctuations of the velocity.

\end{abstract}

\maketitle

\section{\label{sec:level1}Introduction}

Monolayer graphene (MLG) is a two-dimensional Carbon layer that has been of increasing interest for to scientific community since  its first experimental realization in 2004 \cite{Novoselov04,Berger04,Hashimoto04}. 
Indeed, its chirality and linear dispersion at low energies are responsible for its fascinating properties \cite{DasSarma11} such as Klein tunneling \cite{Katsnelson2006}, quantum Hall effect \cite{Novoselov2005} and their potential applications in electronic devices, graphene-based nanocomposites, or chemical sensors \cite{Schedin2007,Wu2008,Rafiee2009,Stankovich2006,Huang20,chen19}. However, these applications are severely limited by the absence of a gap.
Hence, the band-gap opening and the control of graphene bilayer become essential for  applications in various electronic devices.
One way to create a gap in graphene is the selective functionalization, which has been used, for example, with hydrogen adsorption on a moir\'e of Graphene-Ir(111) \cite{Jorgensen16}.
A functionalization by an ad-atom (or ad-molecule) covalently bounded to a Carbon atom is a resonant scatterer for conduction states which strongly affect electronic structure and transport properties \cite{Leconte11b,Lherbier12,Roche12,Roche13,Trambly13,Yang18}.
Since graphene is a zero-gap material with a bipartite lattice, such  functionalization states create so-called \emph{midgap states} at the Dirac energy $E_D$.
Bernal bilayer
graphene (BLG) is a system formed by two layers of MLG translated from one to the other. 
One of its advantages is the control of its gap by applying an external gate voltage \cite{Castro07,McCann06,McCann13}, which opens the way to multiple applications for nanodevices \cite{Zhang09,Overweg18,Kurzmann19}. 
On the other hand, the BLG devices can be based on changes in their electrical conductivity, which can be performed with using the influence of substrate \cite{Zhou07}, vacancies,  ad-atoms or ad-molecules adsorbed on the surface of BLG \cite{Leenaerts09,Mapasha12,VanTuan16,Missaoui17,Missaoui18,
Katoch18,VanTuan16,PINTO20,son20}. 
Recently, it has been shown that single- and double-sided fluorination 
affect strongly conductivity, exhibiting insulating and conducting behavior, respectively \cite{son20}.
From a theoretical point of view, the study of transport by semi-classical methods has been well done (see for instance Refs. \cite{DasSarma11,McCann13}). 
This approach is valid when $E_F$ is far enough from the Dirac energy. 
But for $E_F$ close to Dirac energy, abnormal transport due to quantum localization has been predicted for a random distribution of absorbates \cite{Missaoui17} and some very specific cases of selective functionalization \cite{VanTuan16,Missaoui18}. 
These effects are important when the resonant scatterer concentration (defect concentration) is large with respect to the charge carrier concentrations; 
indeed, each resonant scatterer creates one midgap states at Dirac energy $E_D$. 
Since these quantum effects, beyond the semi-classical behavior, are extremely dependent on the type of functionalization, a more systematic theoretical study is still needed to understand current experimental results and stimulate new experimental studies.

\begin{figure}
\includegraphics[width=8cm]{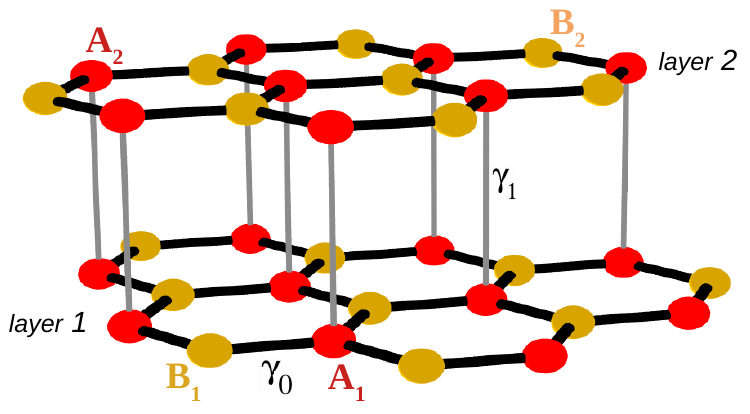} 

\caption{ \label{Fig_bilayer} Sketch of the crystal structure of AB stacked  (Bernal) bilayer graphene (BLG). Atoms A$_1$ and B$_1$ are on the lower layer (layer 1); A$_2$ , B$_2$ on the upper layer (layer 2). 
}
\end{figure}

The unit cell of Bernal BLG contains four Carbon atoms, A$_1$, B$_1$ in layer 1 and A$_2$, B$_2$ in layer 2 (figure\,\ref{Fig_bilayer}). 
Atoms A have three B first neighbors in the same layer and one A neighbor in the other layer, while atoms B have only three A first neighbors in the same layer.
Thus, the local environment of A and B atoms is different, and 
the probability that an atom or molecule will stick to an atom A or an atom B should be different.
It is thus reasonable to think that the functionalization of B atoms is favored.  
This simple argument has been confirmed by DFT calculations \cite{Moaied14} showing that H adsorption energy difference between A site and B site is about $\Delta E = 85$\,meV in favor of B site, when the number of adsorbates is very low. 
For a larger number of adsorbates, one can therefore expect competition between two contradictory effects: on the one hand preferential adsorption on the B-sites of the bilayer, 
and on the other hand adsorption on different sublattices of the same layer as expected in MLG \cite{Boukhvalov08,Moaied14}. 
Indeed in MLG, it exists an interaction between defects states that favors configurations with adsorbates on different sublattices. 
Such  asymmetric adsorption properties between sublattice A and sublattice B have been recently suggested by experimental measurements \cite{Katoch18}, where the distribution of hydrogen adsorbates on the sublattices is adequately controlled. 
Overall, BLG lattice is a bipartite lattice of the two sublattices  $\alpha$ \{A$_1$,B$_2$\} and $\beta$ \{A$_2$,B$_1$\}, from which one expects very specific electronic properties produced by selective functionalization. 
Since BLG is metallic, an isolated functionalization creates an isolated state that is a kind of ``mid-band'' states, so-called midgap states by analogy with MLG. 
In a previous paper \cite{Missaoui18}, we have considered the limiting cases where adsorbates are randomly distributed only on A sublattice or B sublattice of layers 1 while  layer 2 remains pristine.  
On one hand, such a selective functionalization leads to the creation of a gap when sublattice B$_1$ is functionalized. 
This gap is a
fraction of one eV of at least 0.5 eV for a concentration $c$ of adsorbates larger than 1\% of the total number of atoms. 
On the other hand, functionalization of sublattice A$_1$ decreases the effective coupling between layers, and thus the conductivity increases when $c$ increases, since the pristine layer is less perturbed by the disordered layer when $c$ increases. 
These two types of selective functionalization exhibit very different and unusual behaviors. 
This opens the way to the control of electronic properties through selective functionalization, which is experimentally feasible \cite{Katoch18}. 
However, these extreme cases (A$_1$ or B$_1$ functionalization only) 
seem too simple to correspond to the experimental sample. 
Indeed, the complexity of the bipartite BLG lattice requires further theoretical studies of other selective adsorbate distributions.  
This is why it is necessary to study a combined functionalization of several sublattices. 
In particular, we have to consider cases where midgap states are coupled to each other and thus form a midgap band, 
leading to new diffusivity properties that are not a simple combination of the extreme situations studied in Ref. \cite{Missaoui18}, in which midgap states are not coupled together.

In this paper, we present a detailed study of the electronic structure and quantum transport  in BLG with adsorbates (defects) located on two different sublattices among the four sublattices A$_1$, A$_2$, B$_1$ and B$_2$.  
We  analyze how the symmetry is broken between sublattices under this selectivity, which may cause either a gap or abnormal behavior of the conductivity. 
We will pay particular attention to cases where B atoms are preferentially functionalized, 
since these cases should be energetically favorable. 
For instance, under some specific conditions (adsorbates on B$_1$ and B$_2$ sublattices),  a spectacular increase of diffusivity of charge carrier of midgap states band edge is obtained  when the concentration $c$ of adsorbates increases.
The study of conductivity --taking into account all the effects of quantum interference-- requires a distinction between several cases, depending on the value of the inelastic mean free path $L_i$, mainly due to temperature. 
At high temperatures (typically room temperature), we calculate the microscopic conductivity $\sigma_M$; then we will analyze the quantum corrections at low temperature (very large $L_i$ values), i.e. at the localization regime. 
In the latter regime, we also study how localized states due to defects (midgap states), are at the origin of a particular quantum conductivity that cannot be explained by the Boltzmann's transport theory, and which is similar to the one found in quasicrystals \cite{Trambly06,Trambly17}, twisted bilayer graphene \cite{Trambly16} and recently graphene with defects inducing flatbands \cite{Bouzerar20,Bouzerar21}. 

The remainder of this paper is organized as follows.
Section \ref{sec:level2} introduces the model and the formalism to compute the density of states (DOS) and the conductivity. 
Sections \ref{sec:level3} and 
\ref{sec:level4} focus on selective distributions of vacancies distributed in layer 1 only, and the  two layers, respectively. 
Localization effects on conductivity are discussed in Section \ref{sec:level5}. 
Finally, Section \ref{Sec_Conclusion} provides a summary and conclusions.

\section{\label{sec:level2}Electronic structure and numerical methods}

\subsection{TB Hamiltonian}
 The tight binding (TB) Hamiltonian model for BLG with the $p_z$ orbitals only is given by:
\begin{equation}
H = \sum_{(i,j)}t_{ij}|i\rangle\langle j| ,
\label{eq_H}
\end{equation}
where $i$ is the index of $p_z$ orbitals, 
the sum runs over neighbor sites $i$, $j$
and $t_{ij}$ is the hopping element matrix between site $i$ and site $j$. 
In this paper, we consider only the coupling between first neighbors orbitals. 
There are thus two types of coupling 
(Fig.\,\ref{Fig_bilayer}): 
for an intralayer coupling term between first neighboring orbitals A$_1$ and B$_1$ (A$_2$ and B$_2$), $t_{ij}=-\gamma_0=-2.7$\,eV; and for an interlayer coupling term between first neighboring orbitals A$_1$ and A$_2$,  $t_{ij}=\gamma_1=0.34$\,eV 
\cite{Missaoui17}.
For this kind of calculation, 
a more realistic TB model with coupling terms above first neighbors leads qualitatively to similar results \cite{Missaoui17,Missaoui18}. 
We have also checked that such a TB model leads to the results presented here are similar, but the first neighbors TB model allows to better analyze and discuss  the physical mechanisms involved  as it preserves the electron-hole symmetry.
In the Hamiltonian (equation (\ref{eq_H})), the on-site energies are taken equal to zero so that the Dirac energy $E_D$ is therefore equal to zero.    

\subsection{Adsorbate simulation}

We consider that resonant adsorbates are simple atoms or molecules --such as
H, OH, CH$_3$-- that  create a covalent bond with the Carbon atom of the BLG. To simulate this covalent bond,
we assume that the $p_z$ orbital of Carbon, just below the adsorbate, is removed \cite{Pereira08a,Robinson08,Wehling10}. 
In our calculations the vacancies are randomly distributed in two of the four sublattices A$_1$, A$_2$, B$_1$, and B$_2$, with finite concentration $c$ with respect to the total number of atoms. 
Here we study all possible cases of the double type of vacancies:
\begin{itemize}
\item A$_1$B$_1$-Va: Vacancies randomly distributed on sublattices A$_1$ and $B_1$. 
An asymmetric distribution, A$^x_1$B$^{1-x}_1$-Va, where $x$ 
is the proportion of vacancies in the sublattice A$_1$, is also considered.
\item A$_1$A$_2$-Va: Vacancies randomly distributed on sublattices A$_1$ and $A_2$.
\item A$_1$B$_2$-Va: Vacancies randomly distributed on sublattices A$_1$ and $B_2$.
\item B$_1$B$_2$-Va: Vacancies randomly distributed on sublattices B$_1$ and $B_2$.
\end{itemize}
In the following, we call $X$-midgap states the states created by a random distribution of vacant atoms on the $X$ sublattice, with $X = $\,A$_1$, A$_2$, B$_1$, B$_2$, A$_1$B$_1$, A$_1$A$_2$, A$_1$B$_2$, or B$_1$B$_2$.

\subsection{Quantum transport calculation}
\label{Sec_QT}

We used the 
Real Space Kubo-Greenwood (RSKG) method \cite{Mayou88,Mayou95,Roche97,Roche99,Triozon02} which has already been used to study quantum transport in disordered graphene, chemically doped graphene and bilayer (see for instance \cite{Lherbier12,Roche12,Roche13,
Trambly13,Missaoui17,Missaoui18}),
functionalized Carbon nanotubes 
\cite{Latil2004,IshiiInelastic2010,Jemai19}, 
and many other systems (see for instance the recent review Ref. \cite{Fan20} and Refs. therein). 
This numerical method connects the dc-conductivity $\sigma$,
$\sigma = e^2 n D$,
to the density of states $n$ and the diffusion coefficient, 
\begin{equation}
D(E,t)=\frac{\Delta X^2(E,t)}{t}, 
\end{equation}
where the average square spreading $\Delta X^2$
is calculated at every energy $E$ and time $t$ by  using the polynomial expansion method \cite{Mayou88,Mayou95,Roche97,Roche99,Triozon02},
\begin{equation}
\Delta X^2(E,t) = \frac{{\rm Tr} \left( [X,U(t)]^{\dag} \delta(E-H) [X,U(t)] \right)}{{\rm Tr} \,\delta(E-H)},
\label{eq_DX}
\end{equation}
where $U(t)$ is the evolution operator at time $t$, 
$\delta$ is the Dirac function and 
${\rm Tr}$ is the trace. 
This numerical approach has the advantage of using efficiently the  method  in real space. 
It takes into account all quantum effects due to a random distribution of static scatterers in a very large supercell containing more than $10^7$ orbitals.
Here all calculations are done in a super-cell of 1500$\times$1500 cells of Bernal bilayer (4 atoms), with periodic boundary conditions. 
Considering such a huge cell, it is possible to evaluate the traces, ${\rm Tr} A$, in the equation (\ref{eq_DX}) by the average ${\langle A \rangle}$ on a random phase state \cite{Triozon02}. 
Such a calculation may be done by the recursion method (Lanczos algorithm) where the Hamiltonian is written as a tridiagonal matrix in real-space \cite{Pettifor} of dimension $N_r$. 
Here we use $N_r = 1500$ and we checked that presented results do not change significantly when $N_r$ increases. 
Lanczos method,
which has been used in our previous papers \cite{Missaoui17,Missaoui18,Omid20},
leads to a convolution of the DOS by a Lorentzian function which a small width $\epsilon$. 
The DOS is thus obtained by a Lorentzian broadening of the spectrum and 
$\epsilon$ is a kind of energy resolution of the calculation. 
But for systems with a gap, to avoid the tail expansion of the Lorentzian function in the gap, it is more efficient to diagonalize the tridiagonal Hamiltonian of dimension $N_r\times N_r$ 
and to compute the DOS by Gaussian broadening of the spectrum \cite{Lacroix20}.
In the present work a Gaussian broadening is used with the Gaussian standard deviation of 5 meV. 
Note that for energies that are not close to the gap the two methods give almost the same results,
except for small oscillations that look like regular beatings. 
These oscillations are numerical artifacts  depending on 
convergence parameters that we used (see Supplemental Material \cite{SupMat} Sec.\,\ref{Sec_SupMat_gaussian}). They have no effect on the physics discussed here.

The Hamiltonian $H$ (equation (\ref{eq_H})), written in a supercell, takes into account the effects of elastic collisions (static defects, here vacancies). 
Therefore, in the framework of a tight-binding model, all quantum effects --including all multiple-scattering effects-- are  taken into account to calculate the average square spreading $\Delta X^2$ and the diffusive coefficient (equation (2)) without inelastic scattering, i.e. at zero temperature. 
At finite temperature $T$, the inelastic scattering caused by the electron-phonon interactions are implanted by using the approximation of Relaxation Time Approximation (RTA). 
For details of the implementation of the RTA see the appendix of Ref. \cite{Trambly13}.
The conductivity in the $x$-direction is thus given by,
\begin{eqnarray}
\sigma(E_{F},\tau_{i}) &=& e^{2}n(E_{F})D(E_{F},\tau_{i}) , \\
D(E_{F},\tau_{i}) &=& \frac{L_{i}^{2}(E_{F},\tau_{i})}{2\tau_{i}} ,\label{Eq_D_tau}\\
L_{i}^{2}(E_{F},\tau_{i}) &=& \frac{1}{\tau_{i}}\int_{0}^{\infty}\Delta X^{2}(E_F,t)e^{{-t}/{\tau_{i}}}dt, \label{Eq_Li}
\end{eqnarray}
where $E_F$ is the Fermi energy, $\tau_i$ is the inelastic scattering time, $n(E) = {\rm Tr}\, \delta(E-H)$ is the total density of states (total DOS), $D$ the diffusivity  along the $x$-axis, and $L_{i}$ is the inelastic mean free path. $L_{i}(E_F,\tau_{i})$ is the typical distance of propagation during the time interval $\tau_{i}$ for electrons at energy $E$.
$\tau_i$ is the time beyond which the velocity autocorrelation function goes exponentially to zero \cite{Trambly13}.  

$L_i$ is the distance beyond which a wavepacket loses  its phase coherence due to inelastic scattering processes, whereas elastic scattering events do not destroy the phase coherence. 
We know that $L_i$ decreases when the temperature $T$ increases, 
however the exact function of $L_i$ versus $T$ is unknown.
This is why we consider different cases according to different possible values of $L_i$. Indeed,
three different transport regimes may exist depending on $L_i$ value with respect to the elastic mean free path $L_e$, which is the average distance between two elastic scattering events. 
When $L_i \gg L_e$,  multiple scattering effects (such as weak or strong localization) strongly affect the transport and the conductivity is ``macroscopic'' in the sense that it is established over large sample sizes. 
This happens at sufficiently low temperature $T$, 
and then $\sigma$ decreases when $L_i$ increases (i.e. $T$ decreases).
For smaller $L_i$ values, since $L_i > \sim L_e$, i.e. larger temperature, $\sigma(L_i)$ reaches a conductivity plateau close to the maximum $ \sigma$ value, $\sigma_M$, as shown in Sec.\,\ref{sec:level5}. 
This regime is called the diffusive regime, where $\sigma(L_i)$ is almost independent on $L_i$ over a large $L_i$ range depending on the energy $E_F$. 
Examples presented in Sec.\,\ref{sec:level5} show that the conductivity plateau corresponds to $L_i$ values from few nm to few 10\,nm, which may correspond to high temperature and room temperature, respectively.  
In this case, the conductivity of a sample 
depends only on the quantum scattering over small distances which are typically of the order of magnitude of the distances between static defects ($L_e$); this is the reason why we call  $\sigma_M$ the ``microscopic'' conductivity. 
The situation $L_i < L_e$ is an extreme case that one should not often reach in real materials. This corresponds to the case of very pure materials with very few static defects. The conductivity is independent of static defects, and thus $\sigma(L_i)$ increases when $L_i$ increases. 

At each energy, the microscopic diffusivity $D_M$ and microscopic conductivity $\sigma_M$ are defined as the maximum value of $D(\tau_i)$ and $\sigma(\tau_i)$, respectively.
It is also interesting to have an estimate of the $L_e$ values, and the $L_i$ values corresponding to the diffusive regime i.e. $\sigma(L_i) \simeq \sigma_M$. 
We compute  the elastic mean free path $L_{e}$ along the $x$-axis, 
from the usual phenomenological formula 
\cite{Trambly13},
\begin{equation}
\label{le}
L_e(E) = \frac{1}{V_{0}(E)}\, {\rm Max}_{\tau_i} \left\{ \frac{L_i^{2}(E,\tau_i)}{\tau_i} \right\} = \frac{2 D_M(E)}{V_{0}(E)},
\end{equation} 
where the velocity $V_0$ is the slope of $L_i(\tau_i)$ at very small $\tau_i$.
It is important to note that such a definition of $L_e$ is not very accurate, and this calculation can only give an order of magnitude of the average distance  between two elastic scattering events. 
Indeed, the formula (\ref{le}) is not always valid when the electronic structure is strongly modified by static defects. 
Moreover, $V_0$ is overestimated
since the numerical calculations include not only the intraband terms but also the interband terms. In the case of graphene monolayer, we have shown \cite{Trambly16} that these latter increase $V_0$ by a factor of $\sqrt{2}$ which leads to an underestimation of the $L_e$.
However,
roughly speaking, $L_e$ is the $L_i$ value
above which conductivity curve $\sigma(L_i)$ reaches the plateau of diffusive regime due to elastic scattering.
To better define the $L_i$ values corresponding to the diffusive regime, we define the lengths $L_{i1}$ and $L_{i2}$ such as: $\forall L_i \in [ L_{i1} ; L_{i2} ]$, 
$\sigma(L_i) > 0.9 \sigma_M$.
We also determine the value $L_{im}$ such as $\sigma$ is maximum i.e.  $\sigma(L_{im}) = \sigma_M$.
The values of $L_e$, $L_{i1}$, $L_{im}$ and $L_{i2}$ are shown in 
Fig.\,\ref{Fig_LeLi} in the Supplemental Material \cite{SupMat} for different concentrations of the four types of vacancies studied. 
The results show that $L_e \le L_{i1}$ with the same order of magnitude, and the ratio $L_{i2}/L_{i1}$ varies from 5-10 to very large values, depending on the type of defects and  their concentrations.

Microscopic conductivity, which corresponds to the situation where $\sigma(L_i) \simeq \sigma_M$, i.e.  large (or room) temperature limit, is analyzed in Sec.\,\ref{sec:level3} and \ref{sec:level4}.
The $L_i \gg L_e$ limit, i.e. $\sigma(L_i) < \sigma_M$, which corresponds to the
localization regime at low temperature, is analyzed 
Sec.\,\ref{sec:level5}.

\begin{figure*}
\includegraphics[width=7cm]{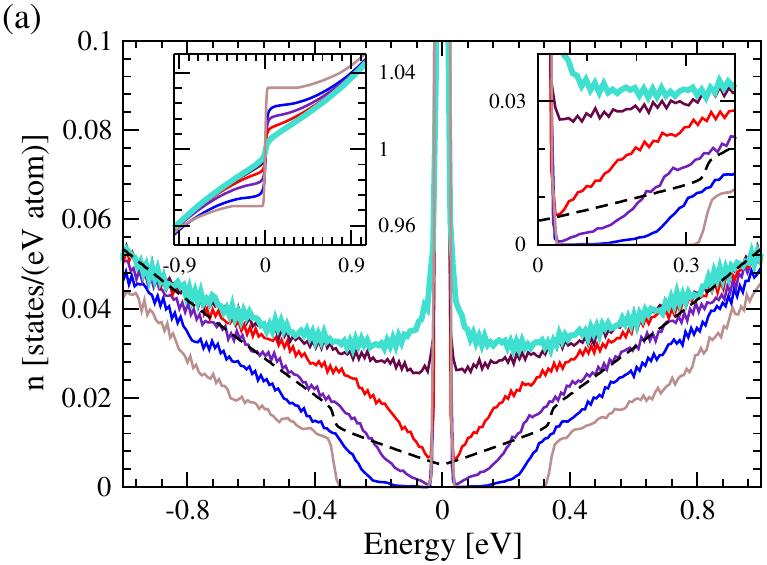} ~~~~
\includegraphics[width=7cm]{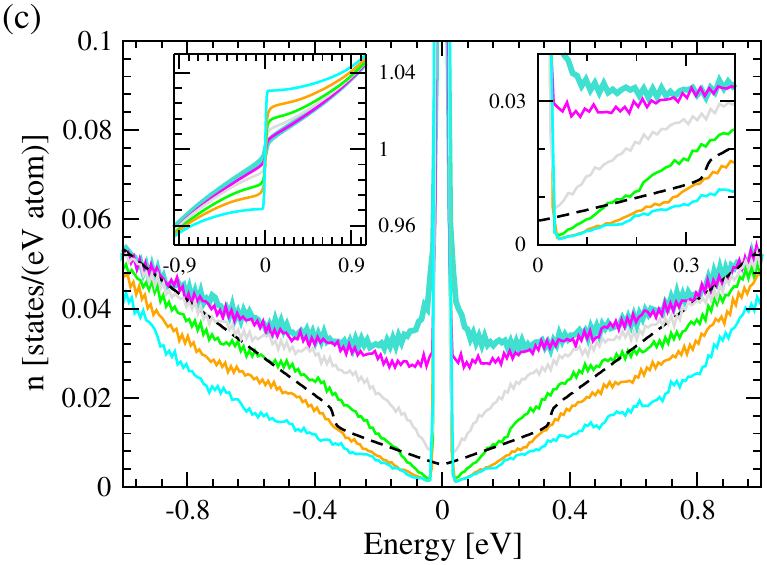}

\vspace{.2cm}
\includegraphics[width=7cm]{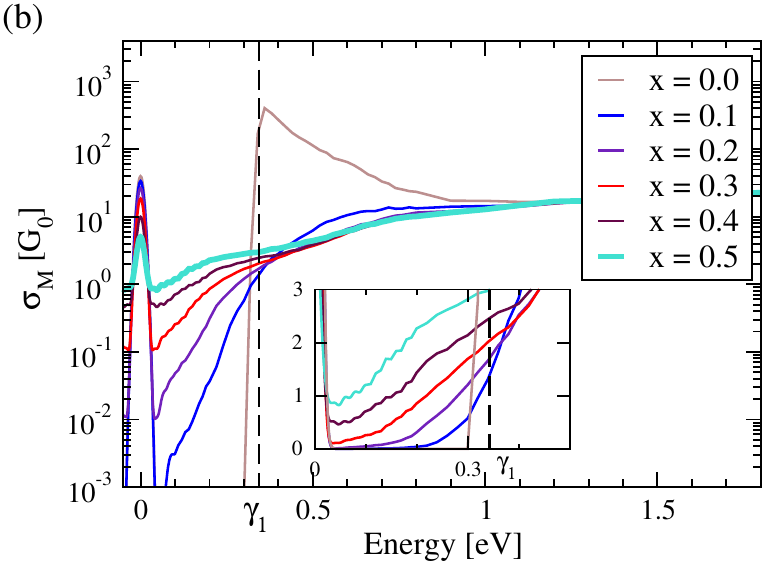}~~~~
\includegraphics[width=7cm]{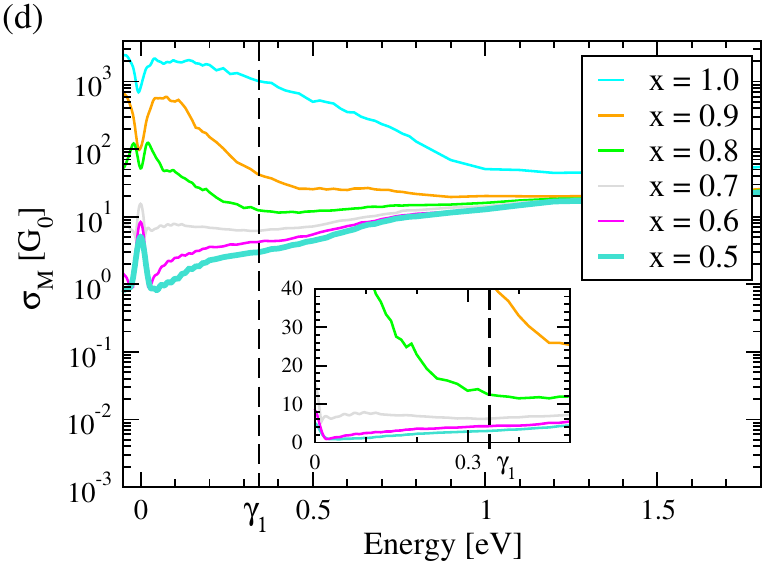}
 
\caption{\label{Fig_A1B1_conf}
BLG with A$^x_1$B$^{1-x}_1$-Va for different distributions $x$ of vacancies between A$_1$ and B$_1$ sites: 
(a-b) $x \in [0;0.5]$ (mainly B$_1$-Va)
and (c-d) $x \in [0.5;1]$ (mainly A$_1$-Va).
(a-c) Density of states $n(E)$, the integrated density of states is represented on  the left insert while the density of states around the Dirac energy E$_D$ is on the right insert.  
(b-d) Microscopic conductivity $\sigma_M(E)$ for the same disorder configurations. 
The total concentration of vacancies is 3$\%$. 
$G_0 = 2e^2/h$.	 
}
\end{figure*}

\section{\label{sec:level3} Vacancies in one layer only}

In this section, we are focusing on the impact of the vacancies distributed on one layer (layer 1) of BLG. 
It should simulate adsorbates or defects that come from the preparation process \cite{Yang18} or induced by the substrate \cite{Ordered}.
For example, in epitaxial graphene on Pt(111) \cite{Ordered}, the authors have shown the appearance of covalent bonds between the Carbon atoms of graphene and the atoms of Pt.
Since the B$_1$ atoms of layer 1 do not have a first neighbor in layer 2, it is likely that their functionalization is favored, 
but the experimental results  \cite{Katoch18} do not show functionalization only on B atoms. 
It is thus important to study an asymmetric functionalization of B$_1$ or A$_1$ sublattice. We first consider a majority functionalization of the B$_1$ atoms (A$_1$ atoms), and we analyze the effect of defect concentrations on a symmetric distribution of vacancies.

\subsection{A$_1$B$_1$-Va asymmetrically distributed}
\label{Sec_A1B1_Assym}

We consider an asymmetric distribution of vacancies: 
A$^x_1$B$^{1-x}_1$-Va, where $x$ ($1-x$) is the proportion of vacancies on sublattice A$_1$ (B$_1$). 
Considering the cases with a total number of vacancies corresponding to a concentration $c = 3\%$ with respect to the total number of atoms,  
the density of states $n(E)$ and the microscopic conductivity $\sigma_M(E)$ are shown in Fig.\,\ref{Fig_A1B1_conf} for different $x$ values.
As presented in Fig.\,\ref{Fig_A1B1_conf_05pc} 
of  the Supplemental Material \cite{SupMat}, the results for $c = 0.5 \%$ show  very similar behaviors.

The different disorder distributions, i.e. value of $x$ between $x=0$ (B$_1$ vacancies only) and $x=1$ (A$_1$ vacancies only), affect strongly the regime around the Dirac energy $E_D$. 
Midgap states at $E_D$ always appeared in both layers.  
Indeed, each A$_1$ missing  orbital of layer 1 produces a A$_1$-midgap state at Dirac energy E$_D$ that spread on B$_1$ sublattice (layer 1) only, and B$_1$ missing orbital produces a B$_2$-midgap states that spread on A$_1$ (layer 1) and B$_2$  (layer 2) sublattices  \cite{Missaoui18}. A$_1$-midgap states and B$_1$-midgap states are coupled by the Hamiltonian and form a band of midgap states with specific  transport properties.
In the extreme cases of vacancies distributed over a single sublattice B$_1$ ($x=0$),
we have shown \cite{Missaoui18} that a gap around the Dirac energy E$_D$ is created. This gap is a consequence of the reduction of the average number of neighbors of atoms in a sublattice.  
For intermediate $x$ values, the gap disappears under the effect of the interactions between midgap states.   
Depending on $x$ values, two scenarios emerge:

(i) For $x \in [0;0.3]$ and $ x \in [0.7;1]$,
the number of A$_1$-midgap states and $B_1$-midgap states are rather different, and many of those states are not coupled to each other and remain isolated with energy $E_D$. The small number of mixed midgap states leads to a small DOS at intermediates energies  
(Fig.\,\ref{Fig_A1B1_conf}(a)).
 
Concerning the conductivity, two  different behaviors are obtained according to the dominant concentration of B$_1$ vacancies $(x \in [0;0.3])$ or  A$_1$ vacancies $(x \in [0.7;1])$. 
The behavior of $\sigma_{M}(E)$ around Dirac energy for $x \in [0;0.3]$  is determined mainly by the effects of the B$_1$ vacancies. For  energies $E$ in the intermediate regime with   $E \leq \gamma_1=0.34$ eV, $\sigma_{M}$ increases when the coupling between midgap states increases, i.e. when A$_1$ and B$_1$ vacancy concentrations are close to each other.
For $x \in [0.7;1]$, results are very sensitive to the concentration of A$_1$ vacancies. $\sigma_M$ increases when $x$ increases. This effect of A$_1$ vacancies affects the microscopic conductivity on a range of energy that does not exceed $1$\,eV as it is shown in Fig.\,\ref{Fig_A1B1_conf}(b). 
In the extreme case $x=1$, a gap appears in the average DOS for the layer with defects (layer 1) \cite{Missaoui18}. It is proportional to the concentration $c$ of vacancies 
and layer 2 behaves more and more like a pristine MLG which gives the ballistic behavior. When $x$ is close to 1, $x \lesssim 1$, the gap in layer 1 disappears, and thus the microscopic conductivity increases when $x$ (close to 1) increases. 

(ii) The interactions between midgap states are important for  $x \in [0.4;0.6]$, and it is maximum for $x=0.5$. Therefore $n(E)$ is larger for energy $E \ne E_D$ (right insert of  Fig.\,\ref{Fig_A1B1_conf}(a)). 
The conductivity behavior is similar to that found in the following section for $x=0.5$. 

\subsection{A$_1$B$_1$-Va symmetrically distributed}
\label{Sec_A1B1_sym}

\begin{figure}
\includegraphics[width=7cm]{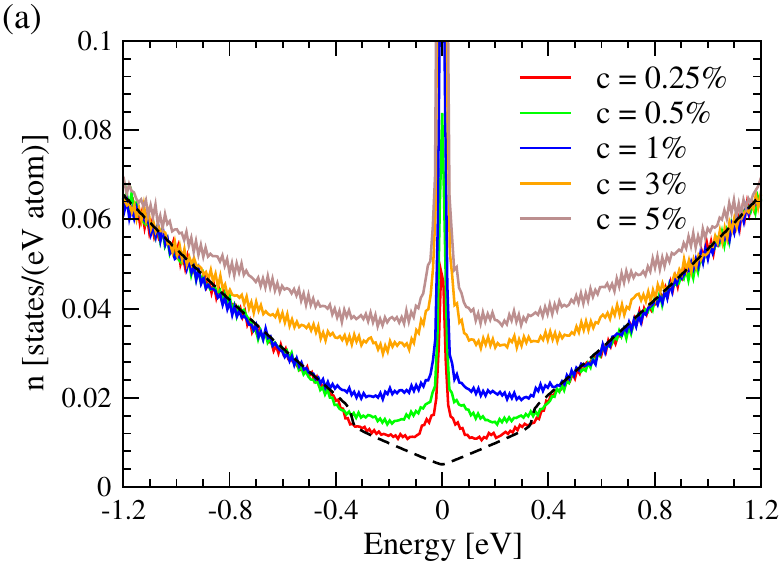} 
\vskip .2cm	
	
\includegraphics[width=7cm]{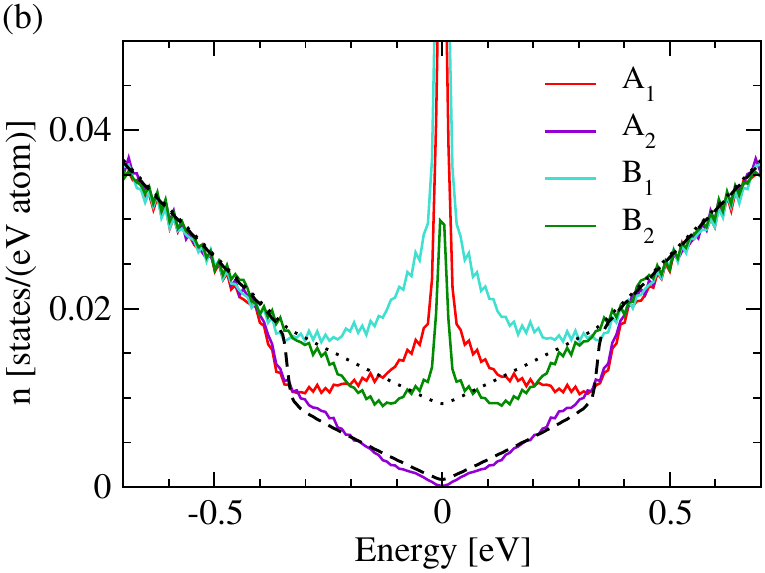}~~ 
	 
\includegraphics[width=7cm]{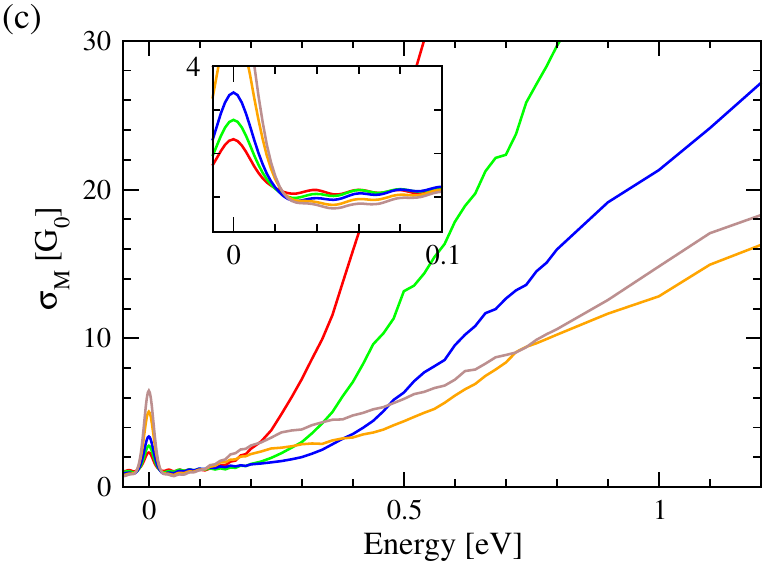} 
\caption{ \label{Fig_A1B1}
Electronic properties in BLG with A$_1$B$_1$ vacant atoms (A$_1$B$_1$-Va), with equal distribution of vacancies between A$_1$ and B$_1$ sublattices:
(a) total DOS (dashed line is the total DOS without vacancies),
(b) average local DOS on A$_1$, B$_1$, A$_2$, B$_2$ atoms  for $c = 0.25$\% 
(dashed line and dot line are LDOS on A and B atom without vacancies),
(c) microscopic conductivity $\sigma_M(E)$.
$c$ is the concentration of vacancies with respect to the total number of atom in BLG. 
$G_0 = 2e^2/h$.
}
\end{figure}

We now study a random distribution of defects equally distributed in sublattice A$_1$ and B$_1$, labeled A$_1$B$_1$-Va. Total DOS $n(E)$, LDOS and microscopic  conductivity $\sigma_M(E)$ are shown in Fig.\,\ref{Fig_A1B1} for several values of vacancy concentrations $c$ with respect to the total number of atoms. 
Since the electron transport through the BLG is mainly determined by the electrons which have energy close to the Dirac point, the conductivity is displayed within a small energy region around the charge neutrality energy $E_D = 0$. 
By inspecting Figs\,\ref{Fig_A1B1}(a-b-c), one can identify several important features. 
(i) For all concentrations $c$ and energy around $E_D$, $0.02\,{\rm eV} < |E-E_D|<0.1$\,eV, $\sigma_M$ presents a minimum plateau at conductivity $\sigma_M \simeq 1.2$\,G$_0$, with $G_0 = 2e^2/h$.  
Thus $\sigma_M \simeq 2 \sigma_M^{mono}$, where
$\sigma_M^{mono} \simeq 0.6$\,G$_0$ is the monolayer graphene (MLG) microscopic conductivity \cite{Yuan10,Gonzalez10,Trambly13,Missaoui17}. 
This shows that the defects affect both planes similarly, although one of the two planes is defect-free.
Moreover, the presence of a plateau almost independently of the concentration, shows that the microscopic quantities in the BLG are not affected directly by interlayer coupling terms, which gives them a behavior similar to MLG.  
This behavior is understandable since the elastic mean free path $L_e$ (see Supplemental Material \cite{SupMat} Figs.\,\ref{Fig_A1B1_Le}) 
is smaller than the traveling distance $l_1$ in a layer between two interlayer hoppings,
$l_1 \simeq 1-2$\,nm \cite{Missaoui17}.
(ii) For energies far from $E_D$, $|E-E_D| > 0.1$\,eV, two behaviors of the conductivity is observed: for $c \leq 2\%$, $\sigma_{M}\simeq \sigma_{B}$, where $\sigma_{B}$ is calculated with the Bloch-Boltzmann approach \cite{Castro09_RevModPhys,McCann13}, and then conductivity is proportional to $1/c$. 
While for $c \geq 2\%$, $\sigma_{M}$ seems to depend less on $c$, and even slightly increases when $c$ increases,  such as for A$_{1}$ vacancies alone or B$_{1}$ vacancies alone \cite{Missaoui18}.

\section{\label{sec:level4} Vacancies in both layers}

In this section we study the combined effect of vacancies distributed in two sublattices that do not belong to the same layer. 
The case  B$_1$B$_2$-Va, which seems to be the most favored case for functionalization, is considered first. These midgap states are coupled to each other and form a midgap band characterized by a very unusual quantum diffusion of charge carriers. 
After, we study the cases of
A$_1$A$_2$-Va and  A$_1$B$_2$-Va, that both produce uncoupled midgap states at energy $E=E_D= 0$.

\subsection{B$_1$B$_2$-Va cases}

\begin{figure}
\includegraphics[width=7cm]{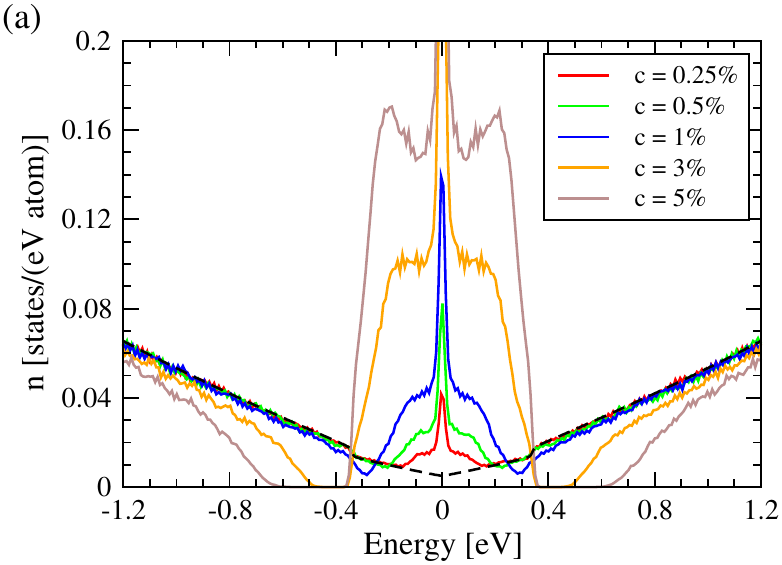} 
\includegraphics[width=7cm]{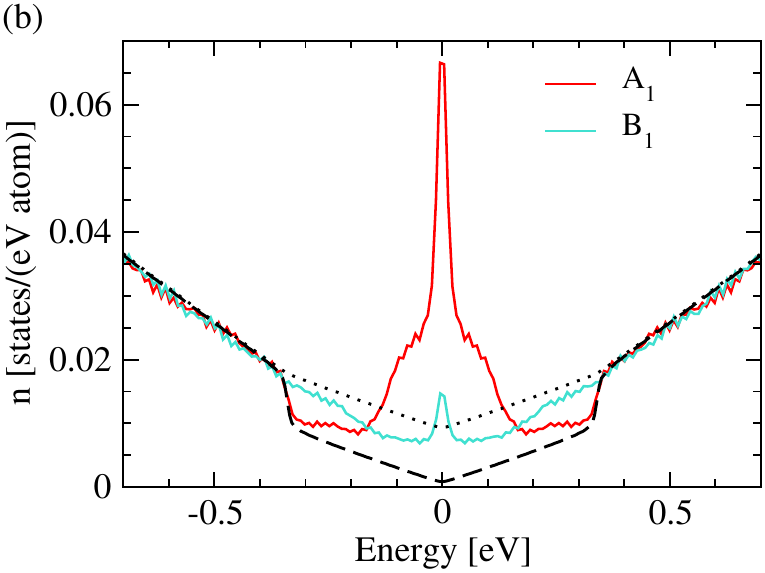} 
\includegraphics[width=7cm]{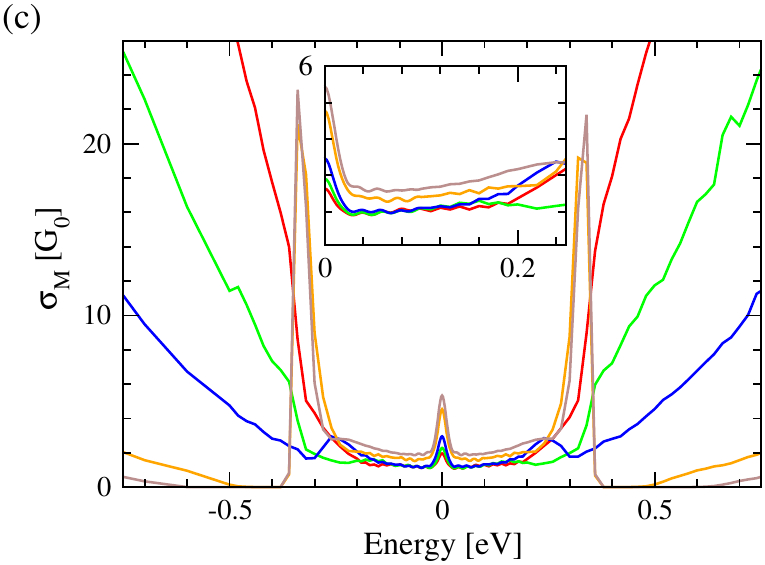} 
\caption{ \label{Fig_B1B2}
(color online)
Electronic properties in BLG with B$_1$B$_2$ vacant atoms:
(a) total DOS (dashed line is the total DOS without vacancies),
(b) average local DOS on A$_1$, B$_1$, A$_2$, B$_2$ atoms  for $c = 0.25$\% 
(dashed line and dot line are LDOS on A and B atom without vacancies). The average local DOS on A$_2$, B$_2$ atoms is obtained by a symmetry with relative atoms A$_1$, B$_1$ respectively.
(c) microscopic conductivity $\sigma_M(E)$. 
$c$ is the concentration of vacancies with respect to the total number of atom in BLG.
$G_0 = 2e^2/h$.	}
\end{figure}

B$_1$- and B$_2$-midgap states are distributed over all the structure with different weights on the atoms A$_1$, A$_2$, B$_1$, and B$_2$ 
(Fig.\,\ref{Fig_B1B2}(b)).
They form a band since B$_1$-Va midgap states and B$_2$-Va midgap states are coupled by the Hamiltonian.  
Their electronic properties are thus very different from those 
of B$_1$ vacancies in BLG  for which a gap proportional to $c$ is formed around $E_D$ \cite{Missaoui18}; while with  B$_1$B$_2$-Va, the B$_1$- and B$_2$-midgap states are coupled, and thus the gap is filled or partially filled by a midgap states band.  
Several regimes are present  depending on both energy $E$ and vacancy concentration $c$. 

For small concentrations $c$, typically $c \le 1\%$, 
there is no gap in the DOS (Fig.\,\ref{Fig_B1B2}(a)) and states around $E_D$ form a narrow midgap states band. 
The corresponding microscopic conductivity $\sigma_M$ presents a plateau (see the insert Fig.\,\ref{Fig_B1B2}(c)) at a value independent on $c$, 
$\sigma_{M}\simeq 2\sigma_M^{mono}$. 

For high concentrations $c$, the density of states (Fig.\,\ref{Fig_B1B2}(a)) around $E_D$ increases significantly, and as a direct consequence, the plateau of conductivity increases $\sigma_{M}> 2\sigma_M^{mono}$.
As explained above (Sec.\,\ref{Sec_A1A2_A1B2}), in each layer the gap due to B-Va increases when $c$ increases, therefore
for large $c$ the midgap states bandwidth becomes smaller than the gap, and the midgap states band becomes isolated from other states by small gaps at $|E| \gtrsim \gamma_1 $ (Fig.\,\ref{Fig_B1B2}(a)). 
The width of this isolated band is $\Delta w \simeq 2\gamma_1$, i.e. $E \in [-\gamma_1,\gamma_1]$.
For large concentrations  $c$, the edge states ($E\simeq\pm \gamma_1$) have a very exotic conductivity $\sigma_M$ which strongly increases when $c$ increases, whereas DOS does not change too much. 
Roughly speaking this spectacular behavior can be explained by considering the coupling between the B$_1$-Va monolayer midgap states and the B$_2$-Va monolayer midgap states.
In monolayer, B-Va midgap states are located on the A sublattice around the B vacancy. 
B-Va midgap states of each layer are not coupled to each other.  
But, since each A orbital are coupled with an A orbital of the other layer, a B$_1$-Va midgap state is coupled with a B$_2$-Va midgap state, with a hopping term $\gamma_{B_1-B_2}$. 
$\gamma_{B_1-B_2} \simeq \gamma_1$, for the smallest $d_{B_1-B_2}$ distance between the B$_1$-Va and the B$_2$-Va (typically first neighbor $B_1$-$B_2$), and $\gamma_{B_1-B_2}$  decreases when $d_{B_1-B_2}$ increases. 
When $c$ increases, the average  $d_{B_1-B_2}$ 
distance decreases and thus the average  $\gamma_{B_1-B_2}$ value increases. As a result, by a kind of percolation  between monolayer B-midgap states of the two layers, the conductivity through the BLG  increases strongly when $c$ increases.  

Finally, the presence of the conductivity plateau for all concentrations (insert Fig.\,\ref{Fig_B1B2}(c)) can 
be understood considering the 
elastic mean free path $L_e$  shown 
in Supplemental Material \cite{SupMat} 
(Figs.\,\ref{Fig_A1B1_Le} and \ref{Fig_LeLi}).
Around $E_D$ energy ($E \in [-0.2;0.2]$\,eV), $L_e<l_1$,
where $l_1 \simeq 1-2$\,nm is the traveling distance between two interlayer hopping events \cite{Missaoui17}. 
Thus, the diffusion of the charge carriers is not affected by the interlayer coupling. The diffusive regime is reached in each layer independently,
and it takes the MLG minimum value  in each layer.
Note that like for other types of vacancies, for energy away from Dirac energy, $|E-E_D| \gg \gamma_1$, Boltzmann behavior is always found.

\subsection{A$_1$A$_2$-Va and A$_1$B$_2$-Va cases}
\label{Sec_A1A2_A1B2}

\begin{figure}
\includegraphics[width=7cm]{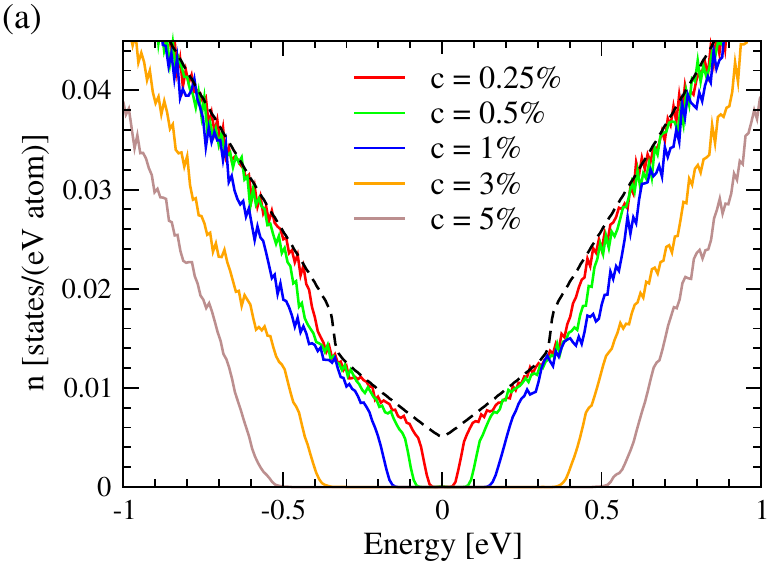} 
\includegraphics[width=7cm]{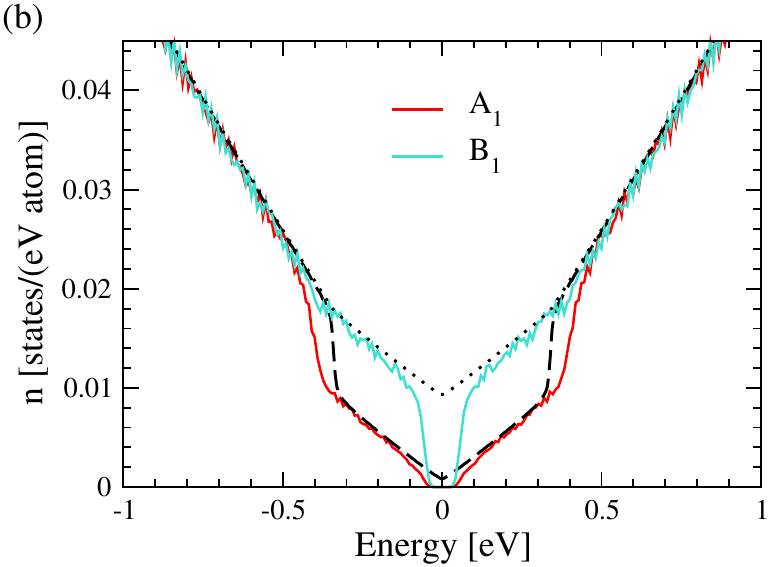} 
\includegraphics[width=7cm]{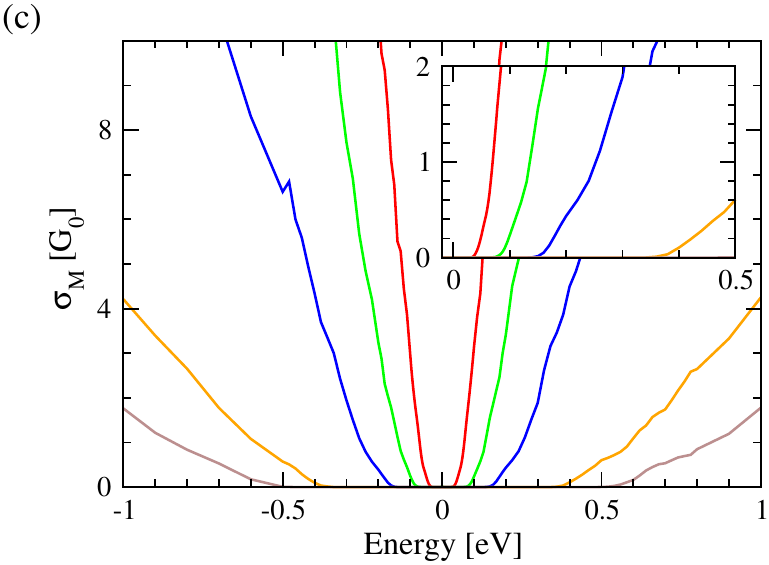} 
\caption{ \label{Fig_A1A2}
(color online)
Electronic properties in BLG with A$_1$A$_2$ vacant atoms:
(a) total DOS (dashed line is the total DOS without vacancy),
(b) average local DOS on A$_1$, B$_1$ atoms  for $c = 0.25$\% 
(dashed line and dot line are LDOS on A and B atom without vacancy). The average local DOS on A$_2$, B$_2$ atoms is obtained by a symmetry with relative atoms A$_1$, B$_1$ respectively.
(c) microscopic conductivity $\sigma_M(E)$.
$c$ is the concentration of vacancies with respect to the total number of atoms in BLG. 
For clarity the midgap states at $E_D=0$ are not shown (see text).
$G_0 = 2e^2/h$.	}
\end{figure}

\begin{figure}
\includegraphics[width=7cm]{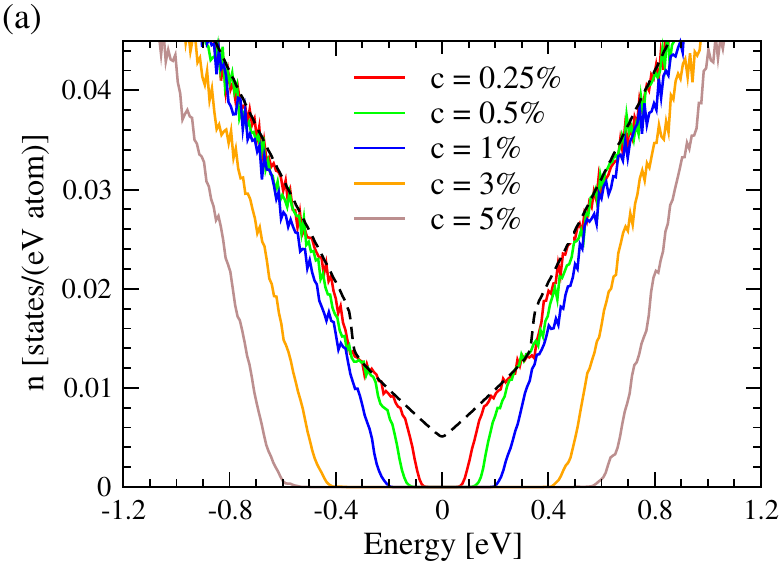} 
\includegraphics[width=7cm]{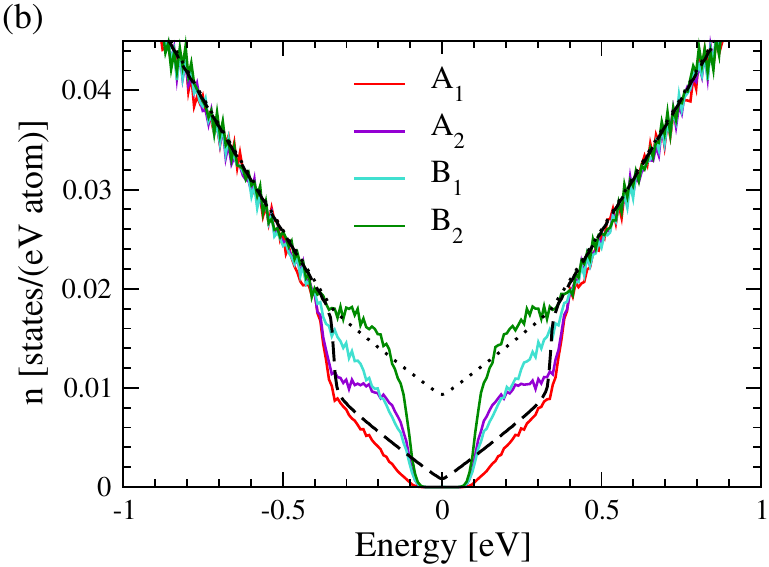} 	
\includegraphics[width=7cm]{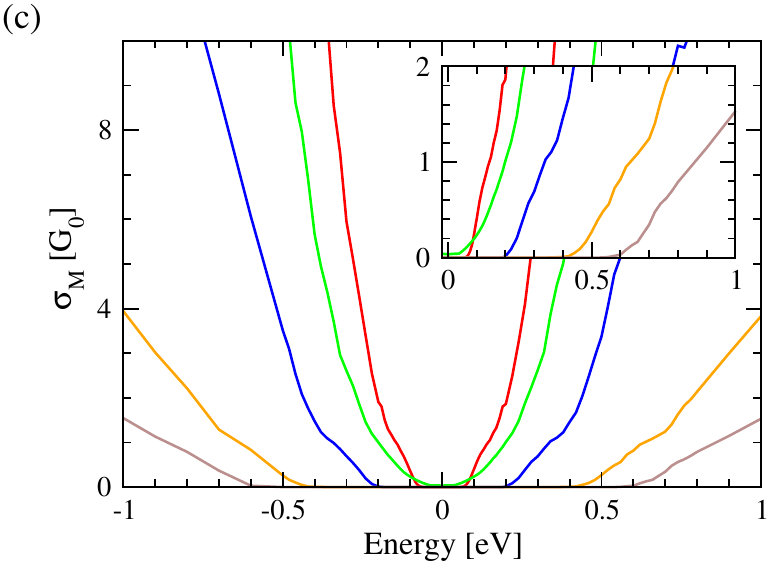} 

\caption{ \label{Fig_A1B2}
(color online)
Electronic properties in BLG with A$_1$B$_2$ vacant atoms:
(a) total DOS (dashed line is the total DOS without vacancy),
(b) average local DOS on A$_1$, B$_1$, A$_2$, B$_2$ atoms  for $c = 0.25$\% 
(dashed line and dot line are LDOS on A and B atom without vacancy),
(c) microscopic conductivity $\sigma_M(E)$.
$c$ is the concentration of vacancies with respect to the total number of atom in BLG. 
For clarity the midgap states at $E_D=0$ are not shown (see text).
$G_0 = 2e^2/h$.}
\end{figure}

The double-type vacancies: A$_1$A$_2$-Va (vacancies randomly distributed on A$_1$ and A$_2$ sublattices) and A$_1$B$_2$-Va (vacancies randomly distributed on A$_1$ and B$_2$ sublattices) are characterized by the absence of coupling between midgap states
and thus all midgap states  remain at energy $E_D=0$.
Indeed, in the case of A$_1$A$_2$-Va, $N$ vacancies on atoms A$_1$ (A$_2$) sublattice produce a set of $N$ uncoupled midgap states at Dirac energy $E_D=0$ that are located on the orbitals B$_1$ (B$_2$) of the same layer \cite{Missaoui18}. As B$_1$ orbitals and B$_2$ orbitals are not directly coupled by the Hamiltonian, midgap states located on B$_1$  and B$_2$ sublattices are not coupled together. 
In the case A$_1$B$_2$-Va, vacancies are vacant atoms of the same sublattice $\alpha$ of the BLG lattice. Corresponding midgap states are thus uncoupled states at $E_D$, located on the  $\beta$ sublattice with a greater weight on the  B$_1$ atoms. 
For clarity these isolated states at $E_{D}=0$ are not shown in the DOSs drawn Figs.\,\ref{Fig_A1A2} and \ref{Fig_A1B2} (see Supplemental Material \cite{SupMat} Sec.\,\ref{Sec_SupMat_gaussian}).  

\begin{figure*}
\includegraphics[width=7cm]{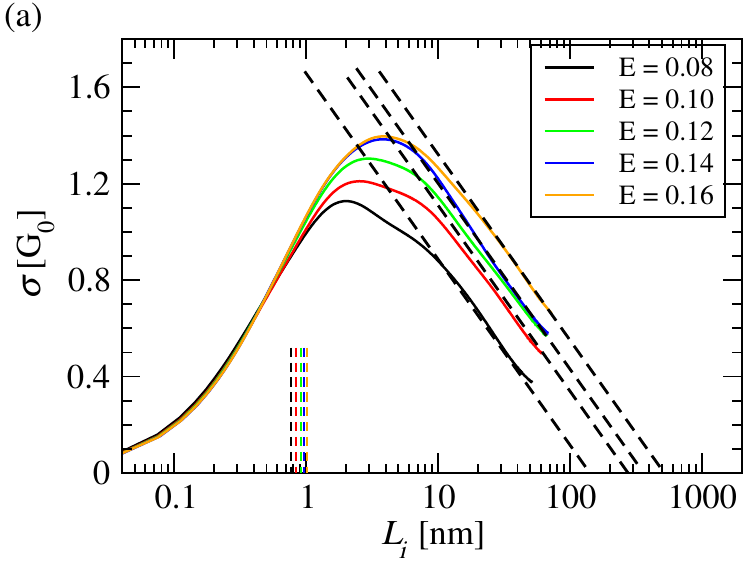}~~~~ \includegraphics[width=7cm]{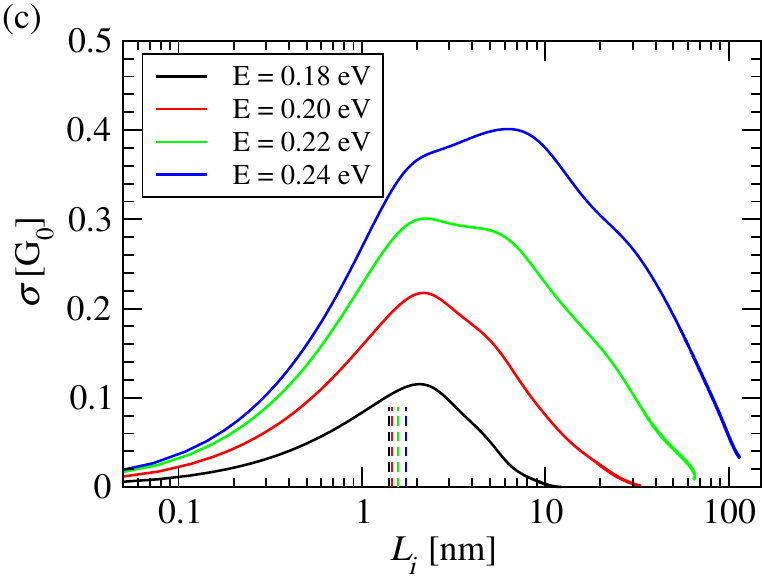}
\includegraphics[width=7cm]{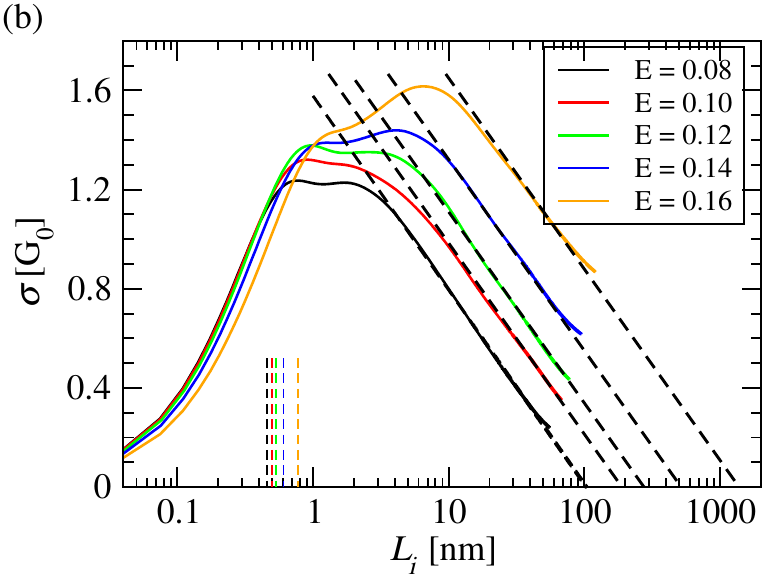} ~~~~~\includegraphics[width=7cm]{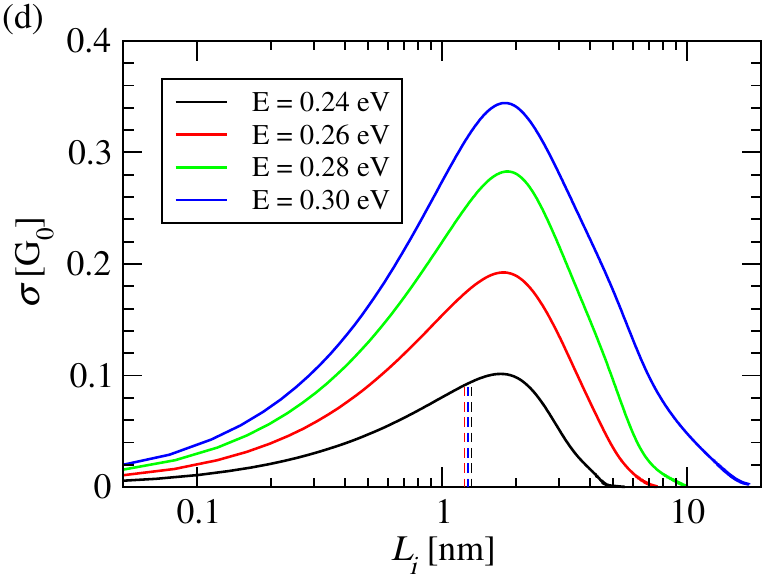}
\caption{ \label{locA1B1} Conductivity $\sigma$ as a function of inelastic scattering length $L_i$ for $c=1$\%. 
(a) Vacancies randomly distributed on atoms A$_{1}$ and B$_{1}$, 
(b) Vacancies randomly distributed on atoms B$_{1}$ and B$_{2}$,
(c) Vacancies randomly distributed on atoms A$_{1}$ and A$_{2}$. (d) Vacancies randomly distributed on atoms A$_{1}$ and B$_{2}$.
$G_0 = 2e^2/h$.
The vertical dashed lines show the value of $L_e$ calculated by Eq. (\ref{le}) for each energy value.  
In panels (a) and (b), the black dashed lines are the extrapolation of $\sigma(L_i)$ curves, using Eq. (\ref{localisation}), to find the localization length $\xi$ at the limit: $\sigma(L_i=\xi)=0$.
}
\end{figure*}

\begin{figure}
\includegraphics[width=7cm]{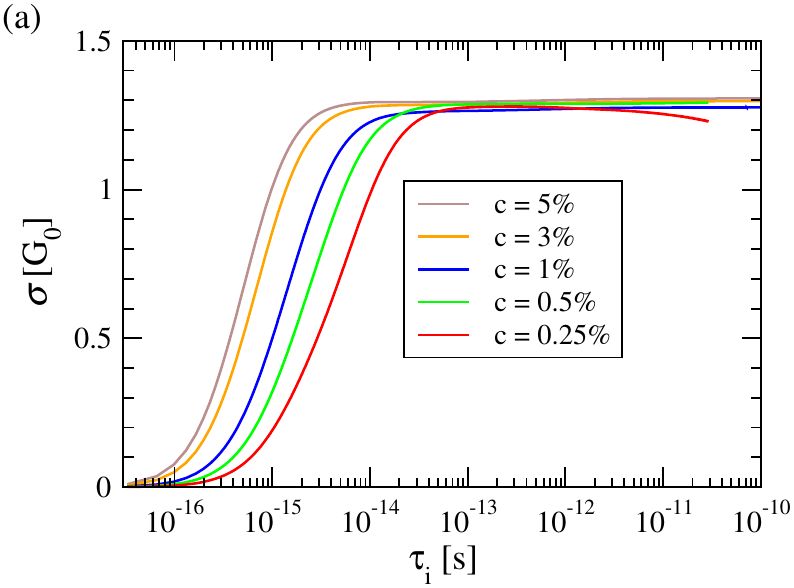} 
\includegraphics[width=7cm]{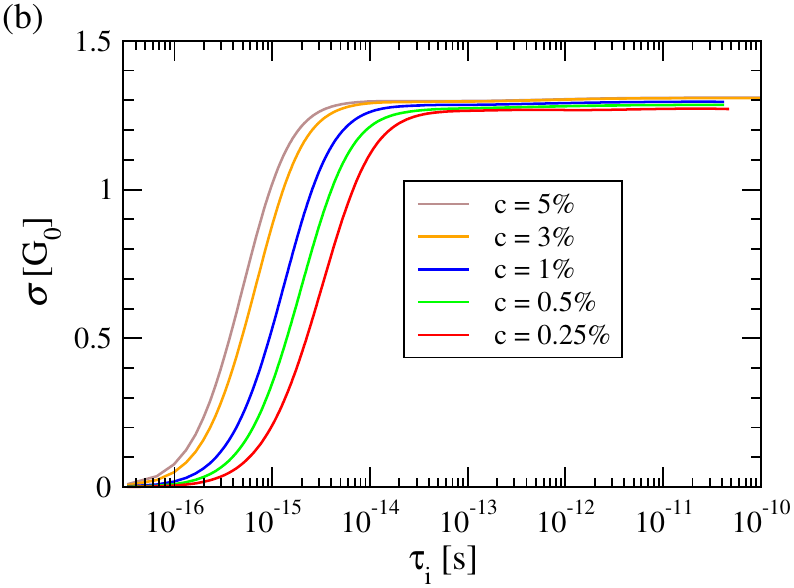}
\caption{ \label{loc_midgap} Conductivity $\sigma(E=E_D=0)$ as a function of inelastic scattering time $\tau_i$. 
(a) Vacancies randomly distributed on atoms A$_{1}$ and A$_{2}$, 
(b) Vacancies randomly distributed on atoms A$_{1}$ and B$_{2}$. 
In both cases midgap states are uncoupled states at $E_D=0$ isolated by gaps. 
$G_0 = 2e^2/h$.}
\end{figure}

In the A$_1$A$_2$-Va case, A$_1$ vacancies and A$_2$ vacancies act on both layers symmetrically and independently because the midgap states of a layer are not coupled with midgap states of the other layer.
Thus, the result is simply the sum of two independent MLG. 
In MLG, vacancies in sublattice A (resp. B) produce midgap states at $E_D$ that are located in sublattice B (resp. A). 
As shown in our previous paper \cite{Missaoui18} by an analysis of the spectrum of bipartite Hamiltonian,
when the concentration $c$ of vacancies increases,  a gap increases around the Dirac energy. 
This gap is a consequence of the reduction of the average number of neighboring atoms of 
sublattice's atoms
which do not contain vacancies. 
Thus,  A$_1$ vacancies (A$_2$ vacancies) create a gap in layer 1 (layer 2) as  
it is clearly shown on the local DOS of atoms A$_1$ and B$_1$ 
(Fig.\,\ref{Fig_A1A2}(b)). 
The total DOS has a gap proportional to the concentration of vacancies $c$ around the Dirac energy $E_D$ (Fig.\,\ref{Fig_A1A2}(a)).

A$_1$B$_2$-Va create also a gap because they are distributed randomly on the same sublattice
$\alpha$ $\{\rm A_1B_2\}$ of BLG.  Total and local DOSs (Fig.\,\ref{Fig_A1B2}(a-b)) confirm the presence of a gap around Dirac energy E$_D$. 

The microscopic conductivity $\sigma_M(E)$ for both types of vacancies A$_1$A$_2$-Va and A$_1$B$_2$-Va  are shown in Figs.\,\ref{Fig_A1A2}(c) and  \ref{Fig_A1B2}(c),  respectively. 
The midgap states at energies $E=E_D$ do not contribute to the conductivity $\sigma_M$ since they are isolated localized states around each vacancy. 
Beyond the gap, $\sigma_M$ decreases when $c$ increases, following a typical Boltzmann behavior \cite{Castro09_RevModPhys}. 

\section{\label{sec:level5} Conductivity versus inelastic scattering
}

In the two previous sections, we have studied the microscopy conductivity $\sigma_M$ which is equal to the maximum value of  $\sigma_M(\tau_i)$ (Sec.\,\ref{Sec_QT}). 
We now consider $\sigma$  versus the inelastic scattering time $\tau_i$ or the inelastic scattering length $L_i$.  
Indeed, the inelastic scattering events, which depend on the temperature, can lead to new behaviors at low temperature due to the multiple scattering i.e. when $L_e \ll L_i$. 
This reveals new quantum effects such as the Anderson localization and the universal conductivity of the midgap states.

\subsection{Anderson localization}

In the framework of the Relaxation Time Approximation (RTA) 
(Sec.\,\ref{Sec_QT}), we compute the inelastic mean free path $L_i(E,\tau_i)$ at every energy $E$ and inelastic scattering times $\tau_i$ 
(Sec.\,\ref{Sec_QT}). 
Figure\,\ref{locA1B1} shows the conductivity $\sigma$  drawn versus $L_i$ for different types of vacancies and different energies close to $E_D$. 
The microscopic conductivity $\sigma_M(E)$ discussed in previous sections (Secs.\,\ref{sec:level3} and \ref{sec:level4}) is the maximum value of the curves $\sigma(L_i)$ at the corresponding energy $E$. 
Each curve $\sigma(L_i)$ has three parts. 
(1) For small $L_i$, typically $L_i \ll L_e$, 
the static defects have no direct effect and $\sigma \propto L_i$. 
This regime is possible at finite temperature only when the defect concentration is extremely low.
(2) For $L_i >  \sim L_e$, $\sigma(L_i)$ reaches a plateau at $\sim \sigma_M$.
For small defect concentrations $c$, this regime can be found for a wide range of $L_i$ values. 
(3) For large $L_i$ values, $L_i \gg L_e$, 
localization regime is reached and $\sigma(L_i)$ decreases when $L_i$ increases.  
In this regime, the so-called quantum corrections, $\Delta \sigma(L_i) = \sigma(L_i) - \sigma_M$, govern the transport properties.

Inelastic scattering collisions are mainly due to electron-phonon interactions, 
and thus $L_i$ decreases when the temperature $T$ increases.
Realistic $L_i$ values are difficult to known, but it is reasonable to consider that
at room temperature and higher temperature, $L_i$ is such as $\sigma(L_i) \simeq \sigma_M$ (plateau regime) and thus the quantum corrections are negligible. 
At low temperatures, i.e. $L_i \gg L_e$, quantum interferences dominate transport properties.

In 2D materials, Anderson localization due to quantum interferences leads to a conductivity varying linearly with $\ln L_i$, \cite{Lee85} and  can be written,
\cite{Trambly11, Trambly13},
\begin{equation}
\label{localisation}
\sigma(E,L_i) = \sigma_0(E) - \alpha G_0 \ln \left(\frac{L_i}{L_e(E)} \right),
\end{equation}
where $G_0 = {2 e^2}/{h}$, 
and $\sigma_0$ values are on the range of $\sigma_M$ values. 
The second term of the right side of equation (\ref{localisation}) is the
quantum correction of the conductivity.
The linear behavior of $\sigma(L_i)$ is clearly seen for cases A$_1$B$_1$-Va and B$_1$B$_2$-Va (Fig.\,\ref{locA1B1}(a-b)).
The fit of the  $\sigma(L_i)$ curve for large $L_i$, gives the $\alpha$ value, $\alpha \simeq 0.34$.
This value is close to the result found in MLG \cite{Trambly13}, BLG with random vacancies \cite{Missaoui17}, 
twisted bilayer graphene \cite{Omid20}, and close too to the prediction of  perturbation theory of 2D Anderson localization \cite{Lee85}, for which $\alpha=1/\pi$.
The localization length $\xi$ can be extracted from the expression (\ref{localisation}) by extrapolation
of $\sigma(L_i)$ curves (Fig.\,\ref{locA1B1}(a-b)) when $\sigma(L_i=\xi)=0$, giving the following expression,
\begin{equation}
\xi(E) = L_e(E) \exp \left( \frac{\sigma_0(E)}{\alpha G_0} \right).
\end{equation}
Since $\alpha$ is a constant, this leads to a simple relationship between $\xi$ and $L_e$, $\xi \simeq 50 L_e$, 
which is between monolayer graphene value with
random vacancy distributions ($13 L_e$) \cite{Trambly13} and that of BLG
in the same case ($13^2 L_e$)  \cite{Missaoui17}.

For A$_1$A$_2$-Va and A$_1$B$_2$-Va cases, 
at energies around the edge of the gap  
(Figs.\,\ref{locA1B1}(c-d)), the decrease of $\sigma(L_i)$ does not follow the equation (\ref{localisation}).
This behavior is more similar to what is generally expected for the conduction by midgap states of graphene \cite{Trambly13}, which are very localized states with abnormal diffusion behavior.\\

\subsection{Universal conductivity of the midgap states}

It is also interesting to focus on the conduction by flatband midgap state themselves i.e., here, midgap states at energy $E_D=0$ that are not coupled to each other by the Hamiltonian (cases A$_1$A$_2$-Va and A$_1$B$_2$-Va cases). 
In these midgap states, the average velocity is zero but conduction is possible due to the inherent quantum fluctuations of the
velocity which are due to the interband contributions of the velocity correlation function \cite{Trambly06,Trambly16,Bouzerar20,Bouzerar21}.
Indeed, in the presence of inelastic scattering these fluctuations are modified \cite{Bouzerar21} and do not cancel completely at large times which allows  electronic diffusion. 
It results a non-Boltzmann conductivity, similar to the one found in quasicrystals \cite{Trambly06,Trambly17}, twisted bilayer graphene at the magic angle \cite{Trambly16}, and graphene with particular defects inducing flatbands \cite{Bouzerar20,Bouzerar21}.
In A$_1$A$_2$-Va and A$_1$B$_2$-Va,  
microscopic conductivity, i.e. small inelastic mean free time $\tau_i$, at midgap-states energy is negligible. But at large $\tau_i$ (large $L_i$), the Kubo-Greenwood conductivity of midgap states is, \cite{Bouzerar21} 
\begin{equation}
\sigma(E,\tau_i) = e^2 n_i(E,\tau_i)D(E,\tau_i),
\end{equation}
where $n_i$ and $D$ are the DOS and the diffusivity (equation (\ref{Eq_D_tau})) in the presence of inelastic scattering. 
Since midgap states are non-dispersive states at $E=0$, isolated by gaps (cases A$_1$A$_2$-Va and A$_1$B$_2$-Va), $n_i$ is the broadening of the Delta function, $c \delta(E)$, by a Lorentzian with a width at half maximum $\eta$, $\eta = \hbar/\tau_i$. Thus at Dirac energy $E_D=0$,
\begin{equation}
\sigma(E=0,\tau_i) = \frac{16}{S} G_0 c \tau_i D(E=0,\tau_i),
\label{Eq_sigma_FTaui_E0}
\end{equation}
where $S$ is the surface of the unit cell.
As shown in Fig.\,\ref{loc_midgap}, for large $\tau_i$, $\sigma(E=0,\tau_i)$ reaches a constant universal value, independant of the defect concentration $c$, which is twice that of graphene \cite{Bouzerar21}: $\sigma(E=0) \simeq 1.3 G_0$.
As shown in Sec.\,\ref{Sec_loc_midgap_A1_B1} of the Supplemental Material \cite{SupMat}, similar behavior is also seen for the midgap states of A$_1$-Va only and B$_1$-Va only (Fig.\,\ref{loc_midgap_A1_B1}).

\section{Conclusion}
\label{Sec_Conclusion}

We have studied numerically the effects on the electronic properties of selective functionalization  distributed over different sublattices
of the Bernal bilayer graphene (BLG). 
We consider the covalent adsorptions of atoms or molecules. 
For Fermi energy $E_F$ far from Dirac energy, typically corresponding to a charge carrier concentration greater than the defect concentration $c$, the adsorbates act as weak scatterers, and the usual semi-classical transport calculations are possible. 
But for smaller doping, typically when the doping is smaller than $c$, $E_F$ is close to Dirac energy the quantum effects  --such as midgap-states or midgap-band, gap, unusual localization-- dominate transport properties. 
Our numerical approach includes all these quantum effects.  

We prove theoretically that controlled functionalization can be an excellent way to tune BLG conductivity. 
This is in agreement with recent experimental results \cite{Katoch18,son20} showing that it is possible to control the functionalization with an adsorbate rate of the order of 1\% of the total number of atoms and to fabricate single and double side adsorbed bilayer graphene. 
We find a wide variety of original behaviors and have classified them according to the functionalized sublattices, the adsorbate concentration $c$, and the energy. 
For example, we give the conditions for opening a mobility gap of several 100\,meV. 
Experimentally,  and according to Ref. \cite{Katoch18}, the Hydrogen adsorption on the B atoms in one layer is energetically favored. For this reason, the study of the specific cases of B$_1$B$_2$-adsorbates is very interesting. An isolated midgap states band is predicted. 
Spectacularly, for $c > \sim 1 \%$, its edge states have a high electrical conductivity due to the large diffusivity of charge carriers, which deserves further investigation.
As the functionalization of atoms can be performed experimentally, one can imagine that those of the B$_1$B$_2$-adsorbates can also be carried out, which makes it possible to control the conductivity. 

The present study contributes to the understanding the electronic properties of localized states --``flatbands''-- due to the combined effect of quantum interferences and geometrical properties (here bipartite lattice).   
This physics of flatbands is currently a major one in condensed matter, either for  field topological insulators or for remarkable electronics (correlation effect, superconductivity) of the moir\'e flatbands in twisted bilayer graphene at magic angle \cite{Cao18a,Cao18b}.  

\section*{Acknowledgments}
The authors wish to thank G.\ Bouzerar, L.\ Magaud, P.\ Mallet, G.\ Jema\"{i}, and J.-Y.\ Veuillen for fruitful discussions. 
Calculations have been performed at the Centre de Calculs (CDC), CY Cergy Paris 
Universit\'e,
and using HPC resources from GENCI-IDRIS (Grant No. 910784).
We thank Y.\ Costes and B.\ Mary, CDC, for computing assistance. This work was supported by the ANR project J2D (ANR-15-CE24-0017) and the Paris//Seine excellence initiative 
(Grant No. 2019-055-C01-A0).

\bibliography{biblio_biAB}
\clearpage

\onecolumngrid
\begin{center}
{\Huge \bf
{Supplemental Material}
}
\end{center}
\vskip 0.5cm

\renewcommand{\thetable}{S\arabic{table}}
\renewcommand{\theHtable}{S\arabic{table}}
\setcounter{table}{0}

\renewcommand{\theequation}{S\arabic{equation}}
\renewcommand{\theHequation}{S\arabic{equation}}
\setcounter{equation}{0}

\renewcommand{\thefigure}{S\arabic{figure}}
\renewcommand{\theHfigure}{S\arabic{figure}}
\setcounter{figure}{0}

\renewcommand{\thesection}{S\arabic{section}}
\renewcommand{\theHsection}{S\arabic{section}}
\setcounter{section}{0}

\noindent
In this Supplemental Material, 
we have shown (Sec. \ref{Sec_SupMat_gaussian})  how the midgap states at energy $E = E_D = 0$ can be removed from the density of states (DOS) and conductivity for clarity. 
In section \ref{Sec_Le}, the calculated elastic mean free path $L_e$ is shown for the different types of studied vacancies.
In section \ref{Sec_diffusiveRegime}, the $L_i$ range corresponding to diffusive regime, i.e., microscopic conductivity, is shown.
In section \ref{Sec_A1B1_asym_c05pc}, the density of states
DOS $n$ and the microscopic conductivity $\sigma_{M}$ for a defect concentration $c=0.5 \%$ of A$_1$B$_1$-Va asymmetrically distributed are presented.
And in section \ref{Sec_loc_midgap_A1_B1}, the conduction by B$_1$-midgap (A$_1$-midgap) states alone is presented.

\section{Numerical treatment  of isolated midgap states at $E_D = 0$}
\label{Sec_SupMat_gaussian}

\noindent
All DOSs are computed from a Gaussian expansion of energies of the spectrum
of the $N_r\times N_r$ tridiagonal recursion
matrix (see main text Secs. II.C).
In these two cases (A$_1$A$_2$-Va and A$_1$B$_2$-Va) where the midgap states at $E_D=0$ are uncoupled, the total DOS $n'$ can  be expressed as the sum of two terms \cite{Missaoui18}:
\begin{equation}
n'(E,\epsilon) = n(E,\epsilon) + x G(E,\epsilon),
\label{Eq_DOS_avec_sans_Lor}
\end{equation}
where $x$ is proportional to the concentration $c$ of vacant atoms, $G$ is the Gaussian function and thus $x G$ is the calculated DOS due to midgap states. $n$ is the DOS without the midgap states. Since isolated midgap states have a zero conductivity, the microscopic conductivity $\sigma_M$ may be expressed as the following manner:
\begin{equation}
\sigma_{M}(E,\epsilon) = e^2 n(E,\epsilon)D_{max}(E)
\label{Eq_Sig_avec_sans_Lor}
\end{equation}

\begin{figure}[h]
\includegraphics[width=7.5cm]{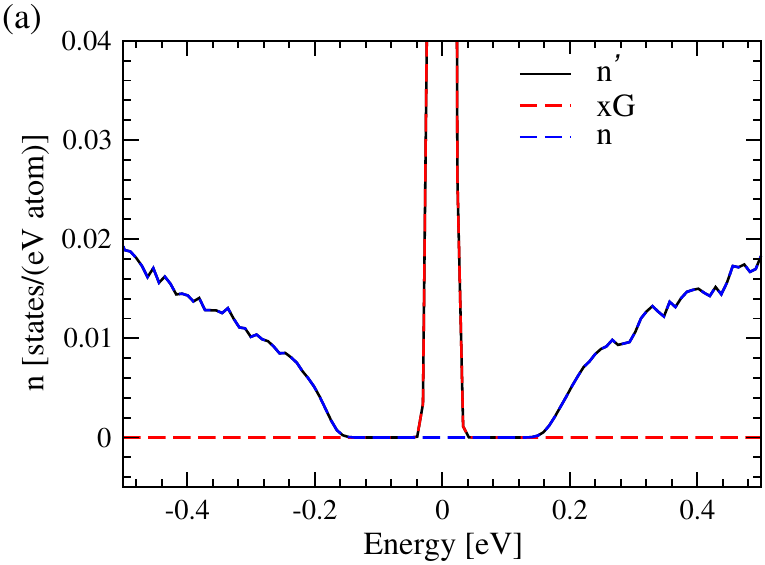} 	
~ \includegraphics[width=7.5cm]{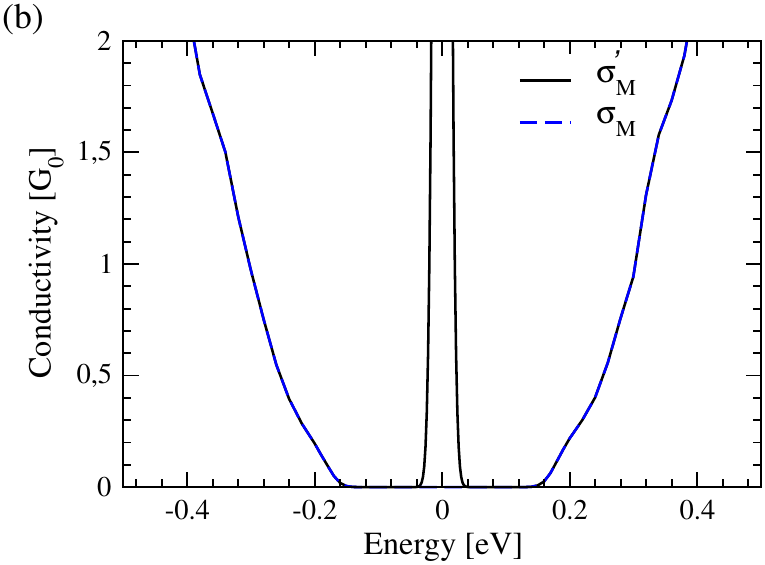} 
	
\caption{ \label{mlor} 
Sketch of equation (\ref{Eq_DOS_avec_sans_Lor})
for $c=1\,\%$ or A$_1$A$_2$-Va: (a) $n'(E,\epsilon)$, $n(E,\epsilon)$, and $xG(E,\epsilon)$ (see Eq. (\ref{Eq_DOS_avec_sans_Lor})) for $\epsilon = 5$\,meV.
(b) Microscopic conductivity (Eq. (\ref{Eq_Sig_avec_sans_Lor}))  with  $\sigma'_m$, and without  $\sigma_M$ the Gaussian term (Eq. (\ref{Eq_DOS_avec_sans_Lor})).  }
\end{figure}

\begin{figure}[]
\includegraphics[width=9cm]{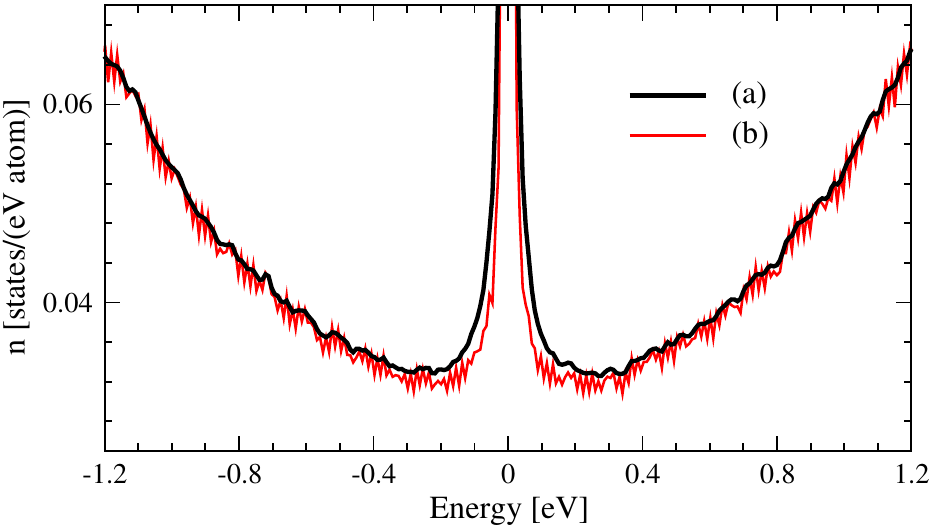} 	
	
\caption{  
\label{Fig_Loren_Gaus} Total DOS for $3\,\%$ of A$_1$B$_1$-Va with  symmetric reparation between the two sublattices: Comparison between (a) Lanczos method (Lorentzian broadening with a half-width at mid-height  of 5\,meV) and (b) Gaussian broadening (Gaussian standard deviation of 5\,meV) after diagonalization of the tridiagonal recursion matrix. See main text Sec. II.C.}
\end{figure}

\noindent
Gaussian broadening of the tridiagonal matrix Hamiltonian gives better accuracy for states around the gap than the Lanczos method, which results in a Lorentzian broadening. 
However, for large concentrations of defects, it induces 
small oscillations that look like regular beatings. 
As shown in figure\,\ref{Fig_Loren_Gaus}, these oscillations are numerical artifacts that are not present when the Lanczos method is used to compute DOS.

\section{Elastic mean-free path}
\label{Sec_Le}

\begin{figure}
	
\includegraphics[width=7cm]{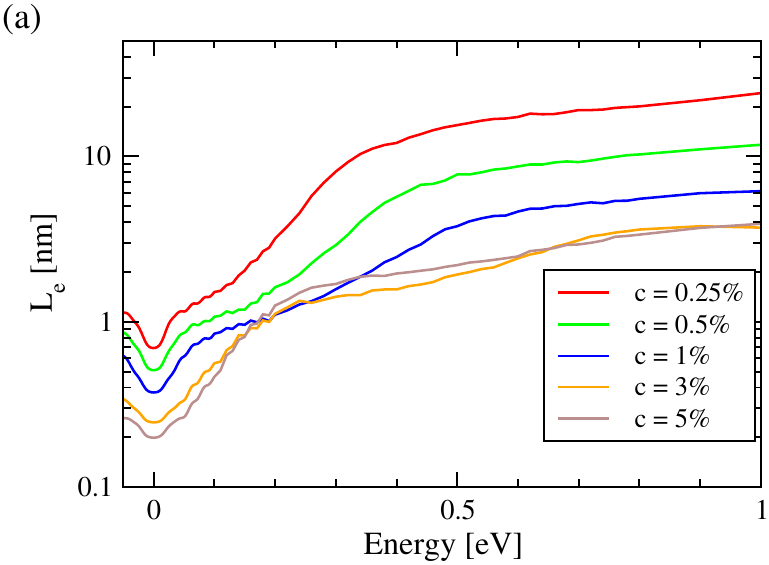} 
\includegraphics[width=7cm]{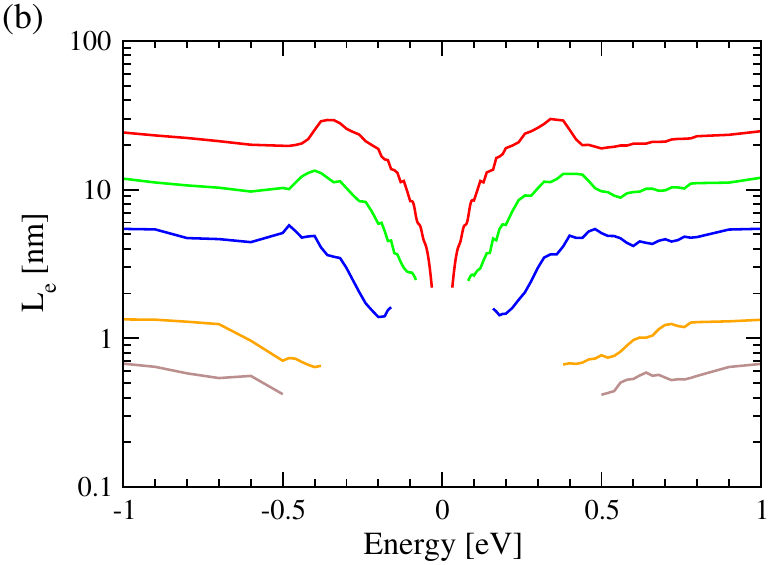} 

\includegraphics[width=7cm]{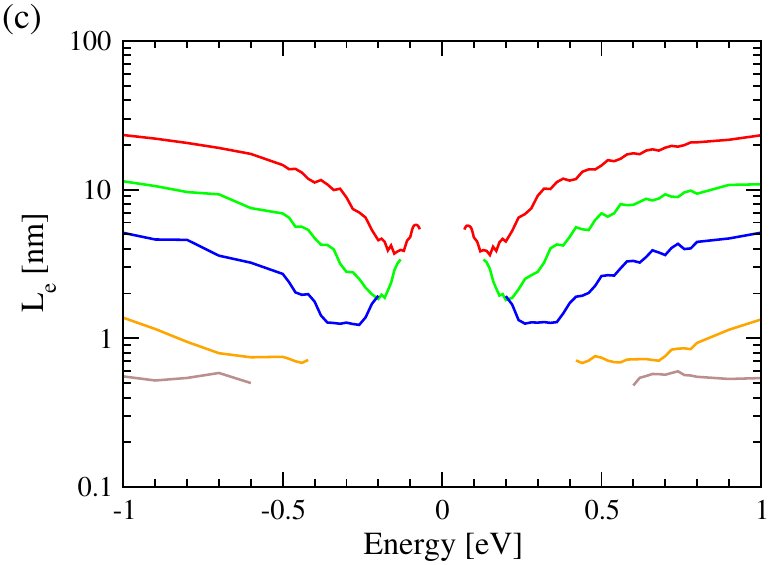}
\includegraphics[width=7cm]{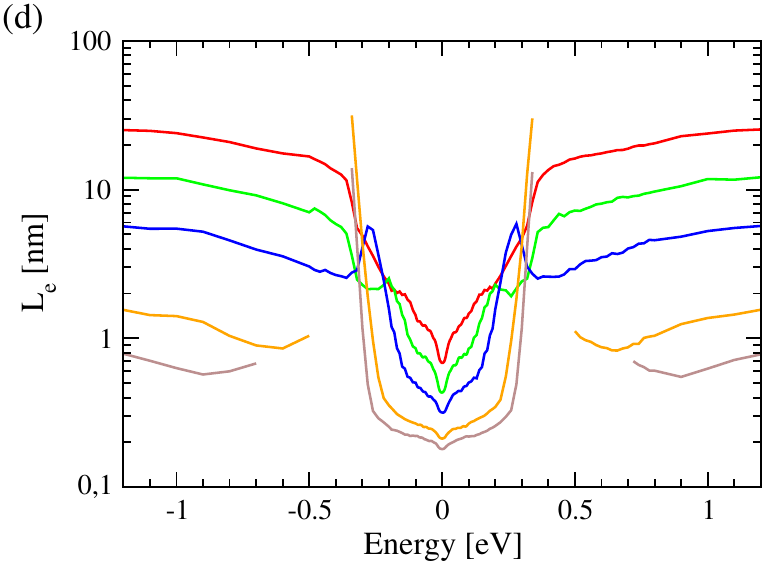}
\caption{\label{Fig_A1B1_Le} Elastic mean free path $L_{e}(E)$ versus energy $E$ for several values of defect concentration $c$: 4 types of vacancies are presented: (a) A$_1$B$_1$-Va, (b) A$_1$A$_2$-Va, (c) A$_1$B$_2$-Va, (d) B$_1$B$_2$-Va.
}
\end{figure}

\noindent
The elastic mean-free path, $L_e(E)$, calculated using equation 
(\ref{le}) 
of the main text is presented in figure\,\ref{Fig_A1B1_Le}. 
Note that $L_e(E)$ is not drawn for $E$ such as the calculated DOS $n(E)$ is very small, because this corresponds to energies in a gap.

\section{Diffusive regime and microscopic conductivity}
\label{Sec_diffusiveRegime}

\begin{figure}[h]
\includegraphics[width=7.5cm]{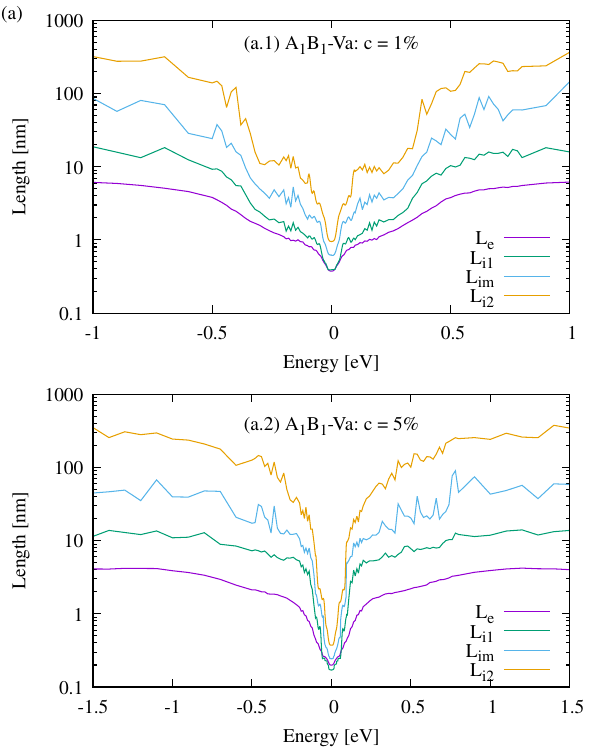} 	
\includegraphics[width=7.5cm]{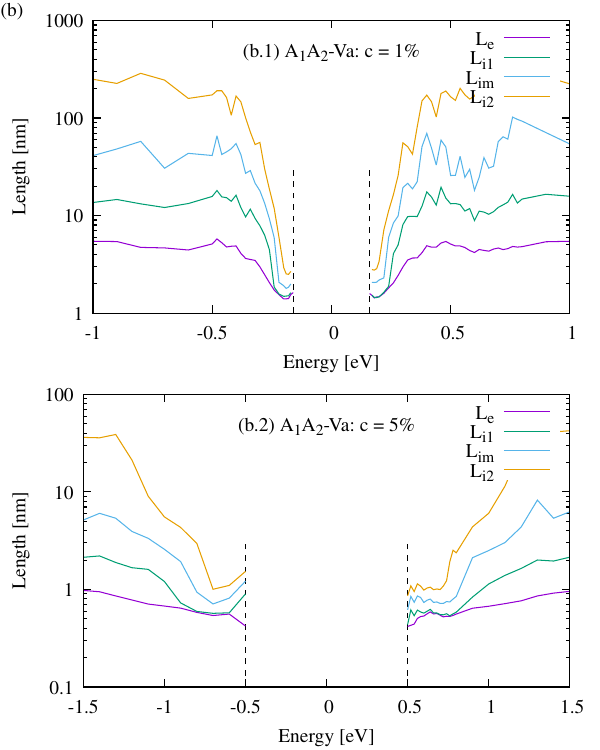} 

\vskip 0.5 cm

\includegraphics[width=7.5cm]{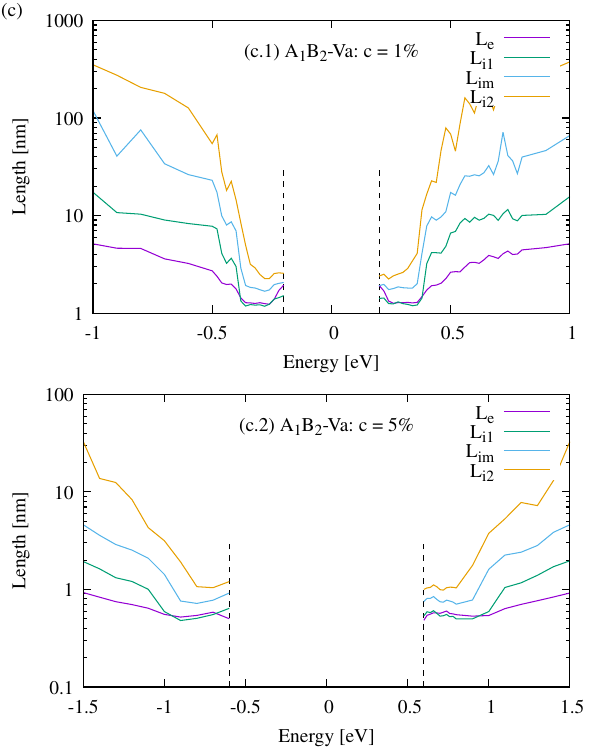} \includegraphics[width=7.5cm]{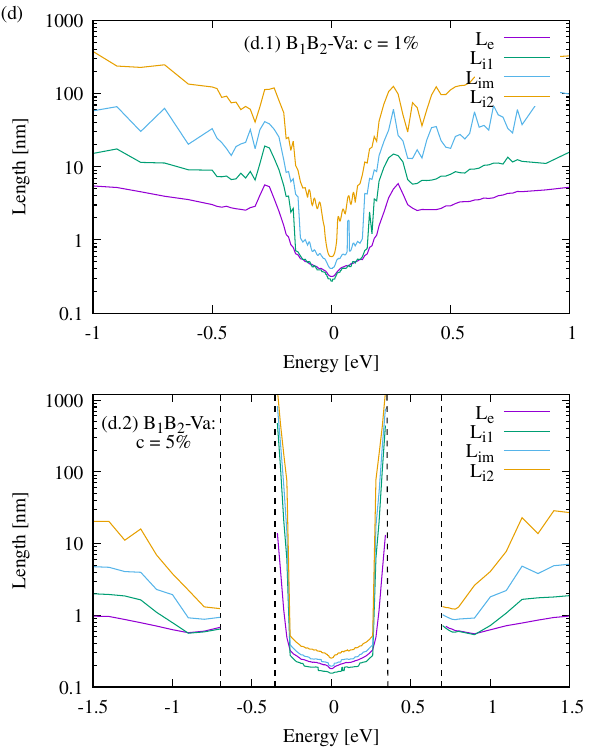}

\caption{ \label{Fig_LeLi} 
Elastic Length $L_e$ and inelastic lengths $L_{i1}$, $L_{im}$, $L_{i2}$, versus energy $E$ for two defect concentrations $c = 1$\% and 5\%, and the four types of vacancies: (a) A$_1$B$_1$-Va, (b) A$_1$A$_2$-Va, (c) A$_1$B$_2$-Va, (d) B$_1$B$_2$-Va. Vertical dashed lines show the gaps.
}
\end{figure}

\noindent
For each energy $E$, the microscopic conductivity $\sigma_M$ is the maximum value of the  conductivity $\sigma(L_i,E)$ which is reached for $L_i = L_{im}$, i.e.  $\sigma(L_{im},E) = \sigma_M(E)$. 
To better define the $L_i$ values corresponding to the diffusive regime, we calculate also the lengths $L_{i1}$ and $L_{i2}$ such as: $\forall L_i \in [ L_{i1} ; L_{i2} ]$, 
$\sigma(L_i) > 0.9\, \sigma_M$.
The values of $L_e$, $L_{i1}$, $L_{im}$ and $L_{i2}$ are shown 
figure\,\ref{Fig_LeLi} for two defect concentrations ($c=1$\% and 5\%) and the four types of vacancies studied. 
The results show that $L_e$ and $L_{i1}$ has the same order of magnitude and $L_e \le L_{i1}$, and the ratio $L_{i2}/L_{i1}$ varies from 5-10 to very large values, depending on the type of defects and  their concentrations.

\section{A$_1$B$_1$-Va asymmetrically distributed with concentration $c=0.5\%$}
\label{Sec_A1B1_asym_c05pc}

\begin{figure}
\includegraphics[width=7cm]{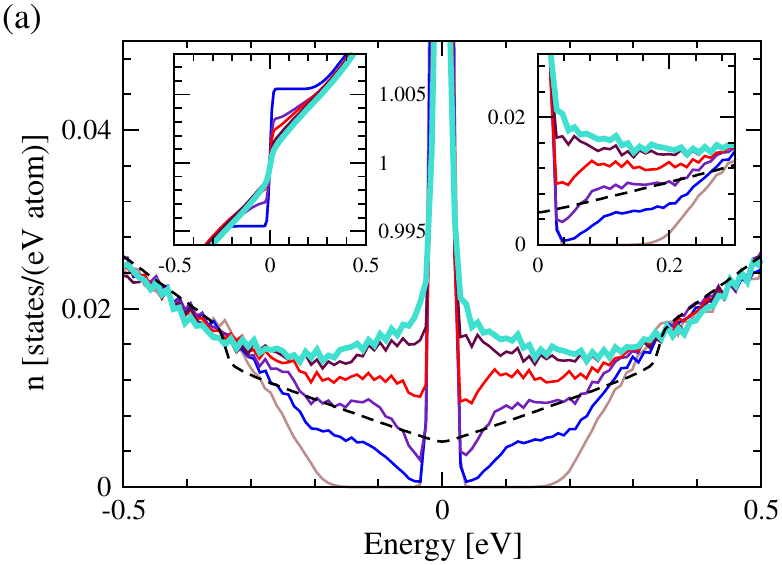} ~~~~
\includegraphics[width=7cm]{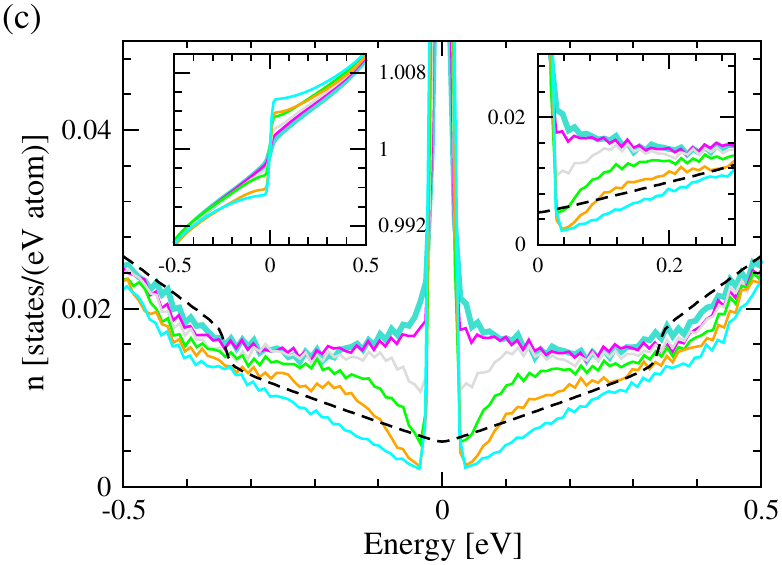}

\includegraphics[width=7cm]{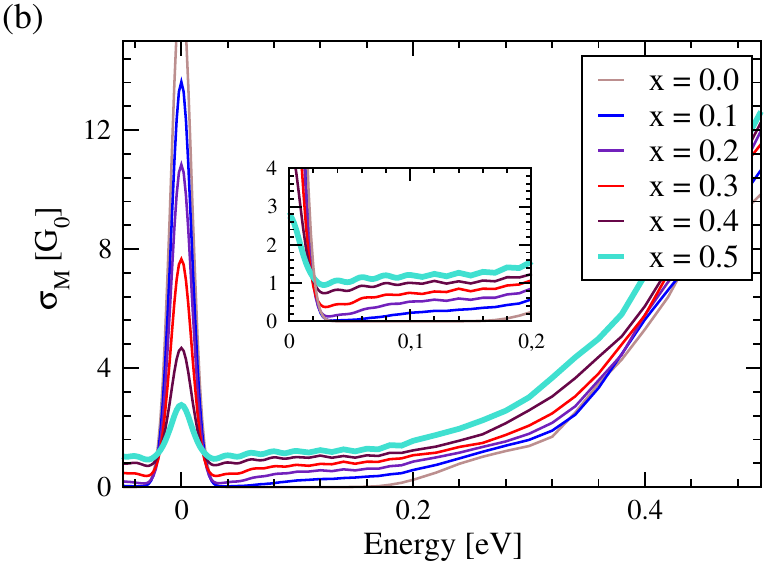}~~~~~ \includegraphics[width=7cm]{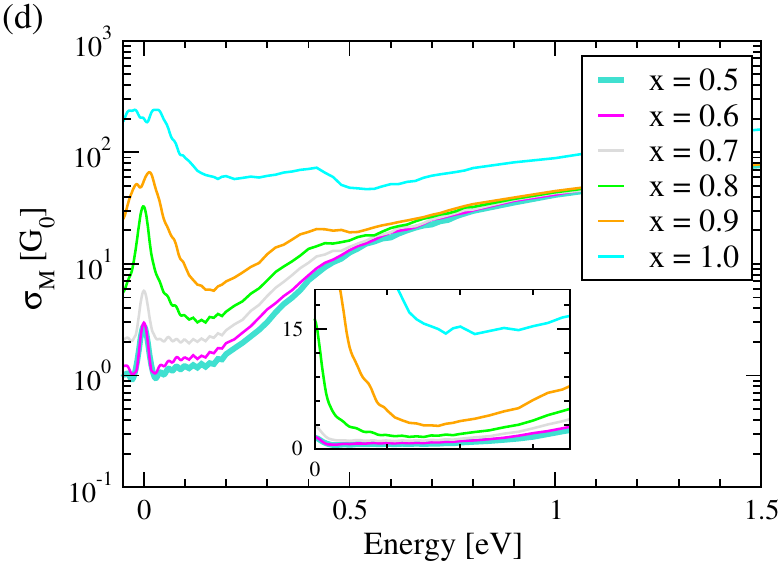}
\caption{\label{Fig_A1B1_conf_05pc}
BLG with A$^x_1$B$^{1-x}_1$-Va for different distributions $x$ of vacancies between A$_1$ and B$_1$ sites: 
(a-b) $x \in [0;0.5]$ (mainly B$_1$-Va)
and (c-d) $x \in [0.5;1]$ (mainly A$_1$-Va).
(a-c) Density of states $n(E)$, the integrated density of states is represented on  the left insert while the density of states around the Dirac energy E$_D$ on the right insert.  
(b-d) The microscopic conductivity $\sigma(E)$ for the same disorder configurations. 
The total concentration of vacancies is $c=0.5 \%$. 
$G_0 = 2e^2/h$.	 
}
\end{figure}

\noindent
The case of A$_1$B$_1$-Va with an asymmetric distribution of vacancies between A$_1$ sublattice and B$_1$ sublattice is discussed in Sec. 
\ref{Sec_A1B1_Assym} 
of the main text. 
Figure\,\ref{Fig_A1B1_conf} 
of the main text is for a total number of vacancies corresponding to $c=3\%$. Here, figure\,\ref{Fig_A1B1_conf_05pc}, a similar figure is shown for $c=0.5\%$. 
The behaviors for $c=3 \%$ and $c=0.5 \%$ are 
very similar. 

\section{Conduction by midgap states for B$_1$-Vacancies and A$_1$-Vacancies}
\label{Sec_loc_midgap_A1_B1}

\noindent
In previous work \cite{Missaoui18} we have studied the unusual microscopic conductivity for limiting cases where defects (vacancies) are randomly distributed in B$_1$ sublattice of A$_1$ sublattice, respectively. 
But in this first work we did not analyze the regime for large inelastic scattering time $\tau_i$ (inelastic mean-free path $L_i$) with respect to elastic scattering time $\tau_e$ (elastic mean-free path $L_e$) as we do in Sec. 
\ref{sec:level5} 
of the present paper.  We thus present here some results for these limiting cases. 

\noindent
At energies $E$ not too close to the Dirac energy $E_D$, 
the $\sigma(L_i)$ curves are similar to those obtained for A$_1$A$_2$-Va and A$_1$B$_2$-Va (figure\,\ref{locA1B1}(c,d) 
in the main text), i.e. for cases where midgap states are uncoupled states at $E_D$. 

\begin{figure}[h]
\includegraphics[width=7cm]{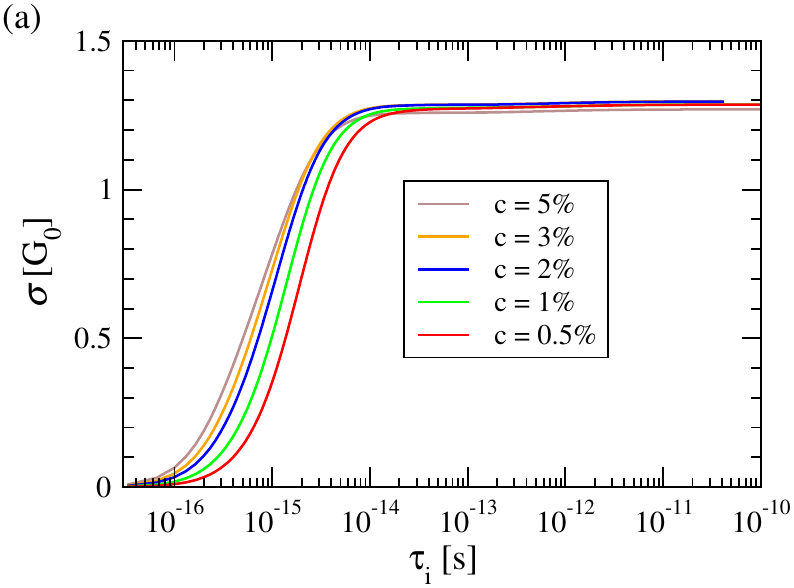} ~~~~
\includegraphics[width=6.7cm]{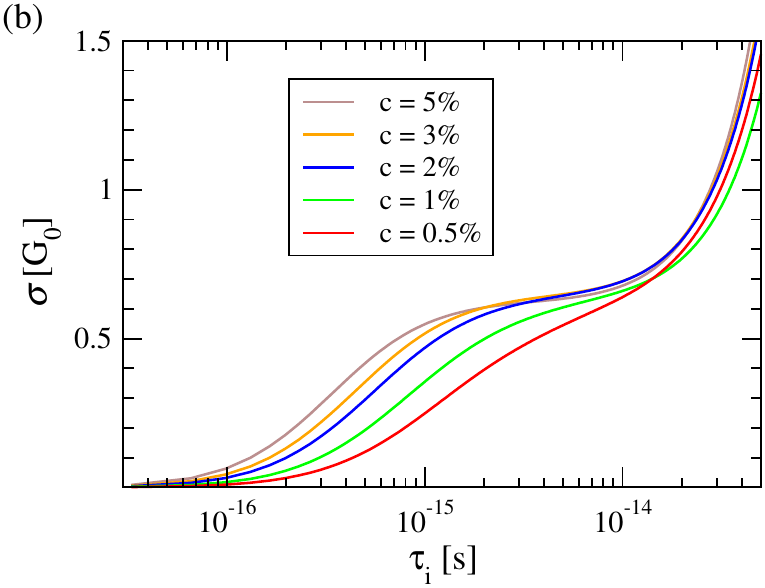}

\caption{ \label{loc_midgap_A1_B1} Conductivity $\sigma(E=E_D=0)$ as a function of inelastic scattering time $\tau_i$ and for different defect concentrations $c$, calculated by the formula
(\ref{Eq_sigma_FTaui_E0}) 
of the main text. (a) Vacancies randomly distributed on the atoms B$_{1}$, (b) Vacancies randomly distributed on the atoms A$_{1}$. In both cases midgap states are uncoupled states at $E_D=0$.
In (a) B$_{1}$-Va case these states are isolated by gaps, 
whereas in (b) A$_1$-Va case they are in the continuum metallic band of the pristine layer (layer 2) \cite{Missaoui18}.
$G_0 = 2e^2/h$.}
\end{figure}

\noindent
At the midgap states energy, $E = E_D=0$, these two limiting cases behave differently from each other (figure\,\ref{loc_midgap_A1_B1}). 
For the B$_1$-Va case 
(figure\,\ref{loc_midgap_A1_B1}(a)), the midgap states at E$_D=0$ are isolated by gaps; and therefore, for large $\tau_i$, $\sigma(E=0,\tau_i)$ reaches a universal constant  value, independent of the defect concentration $c$, which equals two times the graphene one, $\sigma(E=0) \simeq 1.3\,G_0$, as we found for A$_1$A$_2$-Va and A$_1$B$_2$-Va (see Sect. 
\ref{sec:level5} 
and figure\,\ref{loc_midgap} 
in the main text). 
For A$_1$-Va, the situation is completely different 
(figure\,\ref{loc_midgap_A1_B1}(b)), because the midgap states due to A$_1$-Va are located in layer 1 only \cite{Missaoui18}, whereas layer 2 remains pristine.
Therefore, at intermediate $\tau_i$ values, and for sufficiently large defects concentration $c$, the conductivity of the bilayer is driven by the midgap states plateau values of layer 1, $\sigma(E=0,\tau_i) \simeq 0.65 \,G_0$. At large $\tau_i$ values or small concentrations $c$, the conductivity is dominated by the ballistic conductivity through layer 2 and thus $\sigma(E=0,\tau_i) \propto \tau_i^{2}$. 
We believe that this contribution of layer 2 is due to the accumulation of small numerical errors. Yet further investigations are needed as we cannot exclude that this behavior is intrinsic to the system as is observed \cite{Bouzerar21} for the dice model, 
where the peak of localized states is not in a true gap but just at the edge of the continuum. 

\end{document}